\documentclass[a4paper,fleqn,usenatbib]{mnras}

\usepackage{mathptmx}

\usepackage[T1]{fontenc}
\usepackage{ae,aecompl}

\usepackage{graphicx}	
\usepackage{amsmath}	
\usepackage{amssymb}	

\usepackage{booktabs}
\graphicspath{{./figures/}}
\usepackage[acronym, nohypertypes={acronym}]{glossaries}
\usepackage{lmodern}
\usepackage{multirow}
\usepackage[separate-uncertainty=true, multi-part-units=single, range-units=single]{siunitx}
\DeclareSIUnit\parsec{pc}
\DeclareSIUnit\solarmass{M\ensuremath{_{\sun}}}
\DeclareSIUnit\year{yr}
\usepackage{subfig}
\usepackage{threeparttable}

\makeglossaries
\newacronym{dor}{30 Dor-10}{30 Doradus-10}
\newacronym{aca}{ACA}{Atacama Compact Array}
\newacronym{alma}{ALMA}{Atacama Large Millimeter/Submillimeter Array}
\newacronym{aste}{ASTE}{Atacama Submillimetre Telescope Experiment}
\newacronym{atca}{ATCA}{Australia Telescope Compact Array}
\newacronym{bembb}{BEMBB}{broken emissivity modified blackbody}
\newacronym{blast}{BLAST}{balloon-borne large aperture submillimeter telescope}
\newacronym{co}{CO}{carbon monoxide}
\newacronym{casa}{CASA}{Common Astronomy Software Applications}
\newacronym{crtf}{CRTF}{CASA region text file}
\newacronym{dec}{Dec.}{declination}
\newacronym{fov}{FoV}{field of view}
\newacronym{fwhm}{FWHM}{full width at half maximum}
\newacronym{g2d}{G/D}{gas-to-dust}
\newacronym{gmc}{GMC}{giant molecular cloud}
\newacronym{hpbw}{HPBW}{half power beam width}
\newacronym{imf}{IMF}{initial mass function}
\newacronym{ism}{ISM}{interstellar medium}
\newacronym{lmc}{LMC}{Large Magellanic Cloud}
\newacronym{naasc}{NAASC}{North American ALMA Science Center}
\newacronym{pi}{PI}{principal investigator}
\newacronym{ms}{MS}{measurement set}
\newacronym{psf}{PSF}{point spread function}
\newacronym{ra}{R.A.}{right ascension}
\newacronym{rms}{RMS}{root-mean-square}
\newacronym{s/n}{S/N}{signal-to-noise}
\newacronym{sed}{SED}{spectral energy distribution}
\newacronym{sest}{SEST}{Swedish-ESO-Submillimetre-Telescope}
\newacronym{spw}{SPW}{spectral window}
\newacronym{wcs}{WCS}{world coordinate system}
\newacronym{wlm}{WLM}{Wolf-Lundmark-Melotte}
\newacronym{yso}{YSO}{young stellar object}

\title[LMC clump mass function from the ALMA archive]{An ALMA archival study of the clump mass function in the Large Magellanic Cloud}

\author[N. Brunetti \& C. D. Wilson]{
Nathan Brunetti$^{1}$\thanks{E-mail: brunettn@mcmaster.ca} and
Christine D. Wilson$^{1}$
\\
$^{1}$Department of Physics and Astronomy, McMaster University, Hamilton, ON, L8S 4M1, Canada}

\date{Accepted XXX. Received YYY; in original form ZZZ}

\pubyear{2018}

\begin{document}
\label{firstpage}
\pagerange{\pageref{firstpage}--\pageref{lastpage}}
\maketitle

\begin{abstract}
We present \SIlist[list-final-separator={ and }, list-units=single]{1.3;3.2}{\milli\metre} continuum maps of three star forming regions in the \gls{lmc} observed with the \gls{alma}. The data were taken as part of two projects retrieved from the \gls{alma} public archive plus one project observed specifically for this work. We develop a technique to combine maps at these wavelengths to estimate dust-only emission corrected for free-free emission contamination. From these observations we identify \num{32} molecular clumps in the \gls{lmc} and estimate their total mass from their dust emission to range from \SIrange{205}{5740}{\solarmass}. We derive a cumulative clump mass function ($N(\geq M) \propto M^{\alpha+1}$) and fit it with a double power law to find $\alpha_{\mathrm{low}} = -1.76^{+0.04}_{-0.05}$, $\alpha_{\mathrm{high}} = -3.3 \pm 0.2$, and a break mass of $2500^{+300}_{-200}$ M$_{\odot}$. A comparison to the 30 Doradus-10 mass function derived previously from CO (2-1) data reveals a consistent range of clump masses and good agreement between the fitted slopes. We also find that the low mass index of the LMC mass function agrees well with the high mass index for core and clump mass functions from several star forming regions in the Milky Way. This agreement may indicate an extension of the Milky Way power law to higher masses than previously observed.
\end{abstract}




\section{Introduction}
Millimeter observations beyond the Milky Way are beginning to probe the scales of clustered and individual star forming regions known as molecular gas clumps and cores. While filamentary structure has previously been shown to be ubiquitous in Galactic star forming molecular gas \citep{And2010,Arz2011,Kon2015} and in other galaxies \citep[the \acrlong{lmc} in \gls{co}:][]{Fuk2008,Won2011}, it is the higher density clumps and cores within the filaments that directly set the initial conditions for the formation of stars and clusters.

The term ``clump'' is commonly used to refer to molecular gas structures observed with sizes from \SIrange{{\sim}1}{10}{\parsec} and masses up to around \SI{e4}{\solarmass} \citep{Fuj2014} that can go on to form zero to many stars. They are commonly seen as bridging the gap between \glspl{gmc} and cores, and are likely the sites of eventual clustered star formation \citep{Lad1991a,Lad1991b,Lad1992}. Molecular gas cores are typically defined as small (${\lesssim}$\SI{0.1}{\parsec}, ${\lesssim}$\SI{200}{\solarmass}) overdensities where it is expected only one or several stars will soon form \citep{Rei2006b}. Studies of star forming clouds have measured the mass functions of cores and clumps to relate populations of these objects to their local environments and compare to the populations of newly formed stars. The differential mass function is typically characterized by a power law written as
\begin{equation} \label{eq:single_pwr_law_diff}
\frac{dN}{dM} \propto M^{\alpha}
\end{equation}
\citep{Net2009,Kon2015,Fre2017} where $\alpha = -2.35$ is the \citet{Sal1955} stellar \gls{imf} slope. The mass function can also be expressed in cumulative form and through integrating Equation~\ref{eq:single_pwr_law_diff} is fit by a single power law of the form
\begin{equation} \label{eq:single_pwr_law_cum}
N(\geq M) \propto M^{\alpha+1}.
\end{equation}
Other commonly-used functions include a piece-wise double power law \citep{Joh2001,Mot2001,Rei2005,Rei2006a,Rei2006b,Pat2017}, or a lognormal function
\begin{equation}
N(\geq M) \propto 1 - \mathrm{erf}\left[(\ln M - M_{0})/(\sqrt{2}\sigma)\right]
\end{equation}
\citep{Rei2005,Rei2006a,Rei2006b,Net2009,Kon2015}. Galactic measurements of core and clump mass distributions have mostly resulted in power law fits with indices between \num{-1.25} and \num{-2} but with some as steep as \num{-3.2} \citep{Rei2006b,Net2009,Kon2015,Pat2017}. For simplicity, all power law indices in this paper are reported as $\alpha$ from Equations~\ref{eq:single_pwr_law_diff} and \ref{eq:single_pwr_law_cum}.

These functional forms are motivated by the tendency for the mass functions to have a long tail at high masses combined with a flatter portion or turnover at low masses. Physically, the power law tail has been attributed to collapse in molecular clouds where gravity is dominating the motions \citep{Bal2011}, although it has been more directly shown to be occurring due to the appearance of strong density peaks in molecular clouds, however they arise \citep{Tas2010}. Lognormal forms have been argued for as following from the density distribution of supersonic turbulence \citep{Pad1997}, although the presence of supersonic turbulence is not necessary \citep{Tas2010}. The combination of various independent stochastic processes impacting the distribution of molecular gas in clouds has been shown to lead to a lognormal distribution through the central limit theorem \citep{Lar1973,Ada1996}. However, observational evidence for the ubiquity of lognormal distributions may be more difficult to confirm than previously thought \citep{Alv2017}.

In addition to comparing mass distributions between different regions, core mass functions have also been compared to the similarly shaped stellar \gls{imf}. \citet{Cha2003,Cha2005} showed the stellar \gls{imf} follows the original \citet{Sal1955} power law index of \num{-2.35 \pm 0.3} above \SI{1}{\solarmass} with a lognormal turnover at low masses peaking around \SI{0.2}{\solarmass}. This mass distribution has been observed to be universal across environments such as the Galactic disk, young and globular star clusters, and the spheroid or stellar halo \citep{Cha2003,Cha2005,Kru2014}. A similar power law slope has been predicted for both the stellar \gls{imf} and core mass functions \citep{Cha2010,Gus2015}. The observed similarity in slopes has led to the idea that there exists a star formation efficiency that is independent of mass acting to transform the core mass function to the \gls{imf} \citep{Alv2007,And2010,Kon2015}.

At higher masses the comparison is between the mass distributions of entire star clusters and the clump mass functions within their natal \glspl{gmc} \citep{Lad1992}. \citet{Lad2003} compiled a list of \num{76} young star clusters still embedded within a \gls{gmc} and from this derived a cluster mass function with power law index of \num{-2} between ${\sim}$\num{50} and \SI{1000}{\solarmass}. \citet{Fal2012} found the mass functions of star clusters from six different galaxies were all well fit by power laws with indices close to \num{-1.9}.

With the wealth of information gathered on the mass distributions of cores and clumps in the Milky Way, the next step is to extend these studies to other galaxies. This provides new physical environments to test models against as well as larger samples from which to draw statistical conclusions. Two of our nearest neighbors, the Magellanic Clouds, are now well within reach of mm/sub-mm observations of molecular and dust emission for studies of the clump mass function. For example, \cite{Ind2013} derived a \gls{co} mass function of \num{103} clumps at \SI{0.46}{\parsec} resolution in the \glsreset{lmc}\gls{lmc} with the \glsreset{alma}\gls{alma}. While extending beyond the Magellanic Clouds promises a further variety of environments and larger sample sizes, such studies push even the most advanced observatories to their limits. \citet{Rub2015} reported on \num{10} \gls{co} clouds in the \gls{wlm} galaxy with masses of \SIrange{5900}{7.3e4}{\solarmass}. At a distance of \SI{1}{\mega\parsec}, they achieved ${\sim}$\SI{5}{\parsec} resolution with \gls{alma}. More recently, \citet{Sch2017} observed NGC 6822 with \gls{alma} reaching \SI{2}{\parsec} resolution at a distance of \SI{470}{\kilo\parsec}. \num{156} \gls{co} clumps were extracted with masses of \SIrange{9}{3500}{\solarmass}. Thus, if we wish to sample the full range of clump sizes and even start to probe molecular core scales, we are limited to the nearest Local Group members.

The \gls{lmc} is the ideal next step in studying resolved star formation after our Galaxy, given its proximity \citep[\SI{49.97 \pm 1.11}{\kilo\parsec},][]{Pie2013} and nearly face on orientation. It is also a significantly different system from the Milky Way in which to study how stars form, with a lower average metallicity of ${\sim}$1/3 -- 1/4 Z$_{\odot}$ \citep{Rol2002,Duf1984}. Lower metal abundance results in smaller quantities of dust and thus less shielding for star forming regions from UV radiation. This means molecular gas reservoirs may be smaller in mass \citep{Fuk1999,Fuk2001}, limiting the star forming fuel throughout the galaxy. Fewer metals also reduces the cooling through line emission which can change the energy balance in molecular clouds. Finally, the characteristic peak mass of the stellar \gls{imf} has been predicted to shift to higher masses for low metallicity \cite[][albeit for near-zero metallicity conditions]{Bro2005}.

In this paper, we calculate and analyze the dust-derived clump mass function in the \gls{lmc} to facilitate comparison with mass functions of star forming regions in the Milky Way. In \S~\ref{observations} we summarize the observations from each \gls{alma} project we used and we describe our continuum map-making process. In \S~\ref{analysis} we describe our method of isolating the dust-only emission in each region, our clump identification procedure, and the steps we took to fit and characterize our final mass function. In \S~\ref{compare2MW} we discuss our results from \S~\ref{analysis} and place them into the broader context of the study of molecular clump mass functions. The paper is summarized in \S~\ref{conclusion}. Maps of each field included in our mass function are presented in Appendix~\ref{all_maps} as observed emission as well as decomposed into dust-only and free-free-only emission.

\section{Observations and data reduction} \label{observations}
\subsection{Spatial and spectral setups}
We retrieved six publicly available projects from the \gls{alma} archive containing observations of seven fields in the \gls{lmc} covering the star forming regions \gls{dor}, N159W, N159E, N113, N166, GMC 225 and PCC~11546. In addition, we obtained new data in Cycle 4 towards \gls{dor} with the \SI{7}{\metre} \gls{aca}. The projects were observed between 2011 December 31 and 2017 April 21 spanning Cycles \numlist{0;1;2;4}.

For all but one region, multiple pointings were observed across each field in both bands to produce mosaic maps. Mosaic pointings were roughly Nyquist-spaced in a given band to achieve relatively uniform coverage across the inner portion of the maps when imaged together. \SI{12}{\metre} (main array) plus \gls{aca} data were obtained for all fields except N113, which was observed as a single pointing and only with the main array. Numbers of pointings on a field and in a single band range from 1 (N113) to 170 (N166). Mapped areas cover between \num{0.21} square arcminutes (N113) and \num{7.5} square arcminutes (N166; refer to Table~\ref{tab:field_summary3} for areas of each field in both bands).

\begin{figure*}
\centering
\includegraphics{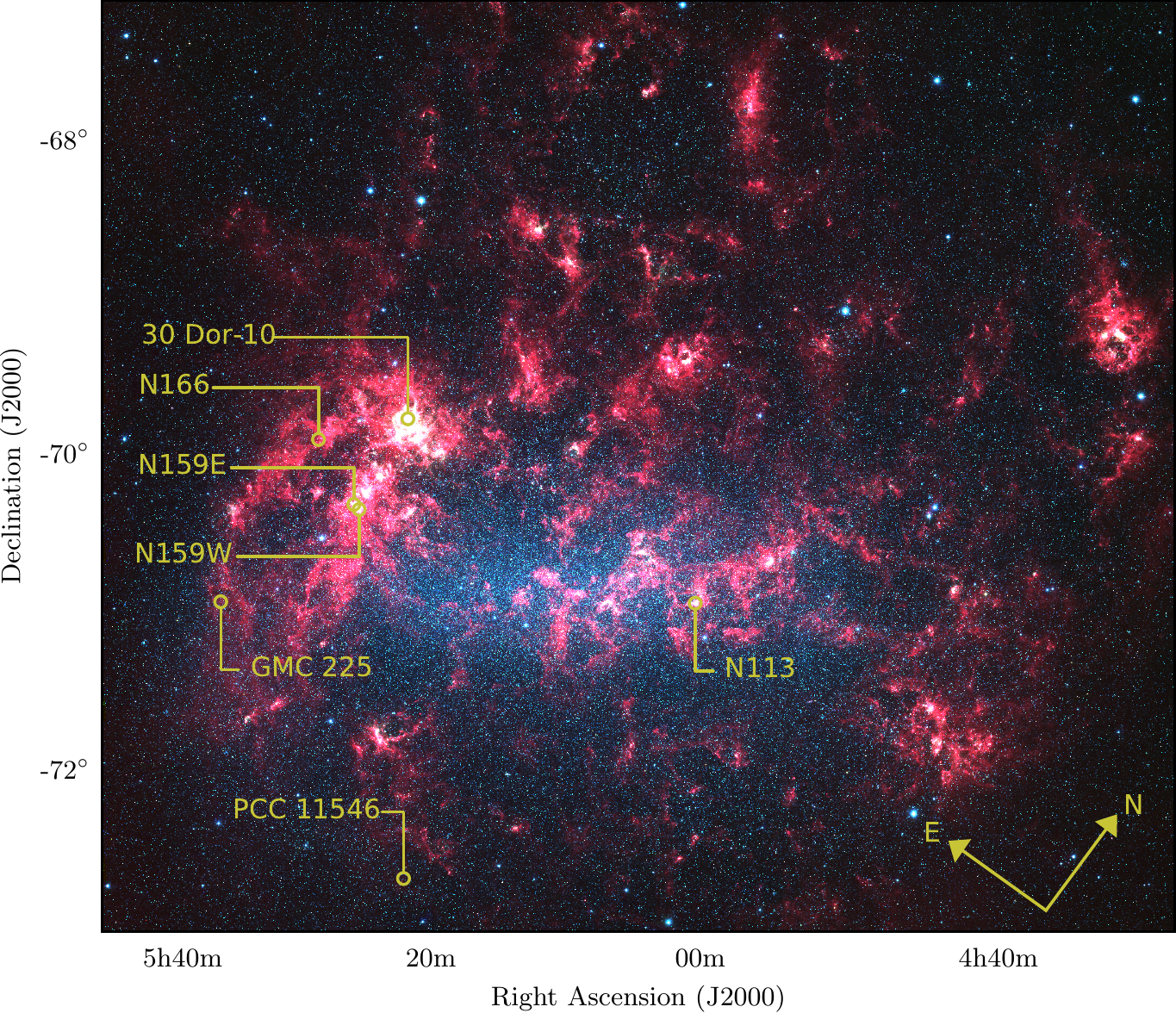}
\caption{SAGE \emph{Spitzer} IRAC three-color image of the \gls{lmc} \citep{Mei2006}. Red, green, and blue are \SIlist[list-final-separator={, and }, list-units=single]{8;4.5;3.6}{\micro\metre}, respectively. The positions of the seven regions discussed in this paper are labeled.}
\label{fig:LMC_fields}
\end{figure*}

To measure dust masses in molecular clumps we focused on continuum observations from \SI{86}{\giga\hertz} to \SI{100}{\giga\hertz} (\SI{{\sim}3.2}{\milli\metre}) and \SI{217}{\giga\hertz} to \SI{233}{\giga\hertz} (\SI{{\sim}1.3}{\milli\metre}) in \gls{alma}'s Bands 3 and 6, respectively. Observations at these frequencies likely also contain significant emission from sources other than thermal dust such as free-free emission. Effective bandwidths used in making the continuum maps ranged between \SI{1.6}{\giga\hertz} and \SI{4}{\giga\hertz}. The locations of the regions analyzed here are marked on the infrared image of the \gls{lmc} in Fig.~\ref{fig:LMC_fields} and a summary of the observational details is given in Table~\ref{tab:field_summary1}.

\begin{table}
\begin{threeparttable}
\caption{\gls{alma} projects used in this paper.}
\label{tab:field_summary1}
\begin{tabular}{@{}c c c c c@{}}
\hline
           & \acrshort{ra}\tnote{a} & \acrshort{dec}\tnote{a} \\
Field Name & (J2000)                & (J2000)                 & Project Code   & Ref.\tnote{b} \\
\hline
\gls{dor}  & 05:38:48               & -69:04:48               & 2011.0.00471.S & 1 \\
''         & ''                     & ''                      & 2016.1.01533.S & 2 \\
N159W      & 05:39:37               & -69:45:48               & 2012.1.00554.S & 3 \\
N159E      & 05:40:09               & -69:44:44               & ''             & 4 \\
N113       & 05:13:18               & -69:22:25               & 2013.1.01136.S & 5 \\
GMC 225    & 05:47:09               & -70:40:16               & 2012.1.00603.S & 6 \\
''         & ''                     & ''                      & 2013.1.01091.S & '' \\
N166       & 05:44:29               & -69:25:43               & 2012.1.00603.S & '' \\
''         & ''                     & ''                      & 2013.1.01091.S & '' \\
PCC 11546  & 05:24:09               & -71:53:37               & 2013.1.00832.S & 7 \\
\hline
\end{tabular}
\begin{tablenotes}
\item [a] Phase center is given for multi-pointing mosaics or pointing center for single-pointing observations (N113).
\item [b] First publication using these data or observing PI name: (1) \citet{Ind2013}; (2) N. Brunetti; (3) \citet{Fuk2015}; (4) \citet{Sai2017}; (5) C. Henkel; (6) A. Kawamura; (7) \citet{Won2017}.
\end{tablenotes}
\end{threeparttable}
\end{table}
\subsection{Background on individual fields}
The Tarantula Nebula, or 30 Doradus, is one of the most actively star forming regions in the \gls{lmc}. It harbors the R136 star cluster which boasts stellar densities between \num{e4} and \SI{e7}{\solarmass\per\cubic\parsec} \citep{Sel2013}. Multiple generations of star formation have occurred in 30 Doradus over the course of ${\sim}$\SI{20}{\mega\year} \citep{DeM2011,Wal1987}. 30 Doradus has been observed as part of several \gls{lmc}-wide surveys targeting different emission sources. In \gls{co}, it has been observed with NANTEN in $^{12}$\gls{co} (1-0) \citep{Fuk2008} and with MOPRA in $^{12}$\gls{co} and $^{13}$\gls{co} (1-0) \citep{Hug2010,Won2011}. The HERITAGE Key Project survey \citep{Mei2013} observed it in dust emission along with the rest of the \gls{lmc}.

The \gls{gmc} \gls{dor} \citep{Joh1998} is part of the Tarantula Nebula within \SI{11}{\parsec} projected distance of R136. \citet{Ind2013} observed \gls{dor} with \gls{alma} in $^{12}$\gls{co} and $^{13}$\gls{co} (2-1) emission as well as \SI{1.3}{\milli\metre} dust continuum at ${\sim}$\SI{1.9}{\arcsecond} resolution. The dust map was used to derive a total H$_{2}$ mass for the \gls{gmc} of \SI{6 \pm 1e4}{\solarmass}. Clumps identified in their $^{12}$\gls{co} and $^{13}$\gls{co} cubes  were used to calculate individual H$_{2}$ masses from both the dust and the \gls{co} data. \citet{Ind2013} calculated a \gls{co} derived mass function for their \num{103} clumps and fit it with a power law of $\alpha = -1.9 \pm 0.2$.

N159 was originally identified by \citet{Hen1956} as an HII region and has since been extensively studied. Noted as the strongest \gls{co} intensity cloud in the initial NANTEN \gls{lmc} survey \citep{Fuk2008}, N159 has been resolved into three major clumps N159W, N159E, and N159S \mbox{\citep{Joh1994}}. Across the region, multi-line excitation analyses on ${\sim}$\mbox{\SI{10}{\parsec}} scales estimated molecular gas kinetic temperatures ranging from \SIrange{10}{25}{\kelvin} \citep{Hei1999,Bol2000}. \gls{alma} observations of N159W  suggest that its constituent group of compact HII regions may have formed through a cloud-cloud collision \citep{Fuk2015}. By adding a dynamical estimate of one of the \glspl{yso} of \SI{e4}{\year}, \citet{Fuk2015} showed that the collisional triggering of star formation likely occurred very recently. N159E contains multiple developed HII regions, the most prominent of which is the Papillon Nebula \citep{Miz2010}. \citet{Sai2017} suggested a three cloud collision is occurring in N159E, with the Papillon Nebula protostar in the overlap region. They also observe a molecular hole around the protostar filled with \SI{98}{\giga\hertz} free-free emission indicating the protostar has recently begun disrupting the molecular cloud from which it formed. N159S does not exhibit any current star formation.

N113 was also identified by \citet{Hen1956} as an HII region. As part of a sample of four molecular clouds (N159W, N113, N44BC, and N214DE) observed with the \gls{sest}, it exhibited significantly lower gas-phase C$^{18}$O/C$^{17}$O abundances compared to molecular clouds in the Milky Way and the centers of starbursts, with a mean across clouds of \num{1.6 \pm 0.3} \citep{Hei1998}. Subsequent observations have shown it to contain the most intense maser in the Magellanic Clouds \citep{Ima2013}, clumpy molecular gas currently forming stars \citep{Sea2012}, and a host of \glspl{yso} as identified by Herschel and Spitzer \citep{War2016}. N113 is related to three young stellar clusters \citep[NGC 1874, NGC 1876, and NGC 1877;][]{Bic1992}. It also contains a rich assortment of molecular species \citep{Par2014}.

N166 is yet another region originally identified by \citet{Hen1956}, and \citet{Fuk2008} identified it as only containing HII regions. Two molecular clouds were observed overlapping its position in the second NANTEN survey and a follow-up observation with \gls{alma} targeted a position between those two NANTEN-observed clouds. N166 was also observed with \gls{aste} in $^{12}$\gls{co} (3-2) revealing densities of \SIrange{e2}{e3}{\per\cubic\centi\metre} and kinetic temperatures between \num{25} and \SI{150}{\kelvin} in a \SI{22}{\arcsecond} beam \citep{Par2016}.

GMC 225 is one of \num{272} \glspl{gmc} identified by \citet{Fuk2008} in the second NANTEN survey of the \gls{lmc} in $^{12}$\gls{co} (1-0) and shows no sign of massive star formation \citep{Kaw2009}. With a \gls{co}-derived total mass of \SI{e6}{\solarmass} and radius of ${\sim}$\SI{73}{\parsec} it was observed as part of a follow up \gls{alma} project. It is ${\sim}$\SI{1500}{\parsec} south of \gls{dor} and east of the molecular ridge.

PCC~11546 is an extremely cold (${\lesssim}$\SI{15}{\kelvin}) dust source identified in the southern limits of the \gls{lmc} as part of the Planck Galactic Cold Cloud catalog \citep{Pla2016}. It also exhibits strong \gls{co} (1-0) emission in the Planck integrated \gls{co} map and the MAGMA \gls{lmc} \gls{co} survey \citep{Won2011}. There appears to be a lack of massive star formation within PCC~11546 and it contains lower density gas than clouds closer to the center of the \gls{lmc} \citep{Won2017}.
\subsection{Calibration}
The visiblity data were retrieved from the archive and calibrated. All fields except \gls{dor} had only raw data available in the archive. \gls{dor} had calibrated visibilities available and we used those data after recalculating the weights based upon the scatter in the visibilities using the \gls{casa} \texttt{statwt} command.

The observations with raw data available were either ``manually'' or pipeline calibrated at the observatory using \gls{casa} \citep{McM2007}. For manually calibrated data we ran the full calibration procedure using the observatory scripts in the latest version of \gls{casa} available at the time, 4.7.2-REL (r39762). Minor editing of the calibration scripts was necessary to account for task and parameter changes from the older versions of \gls{casa}. Edits were also made to ensure the visibility weights were properly calculated throughout the entire calibration procedure.

For pipeline calibrated data, we used the \gls{casa} and pipeline versions closest to those used by the observatory for the original calibration\footnote{All versions of \gls{casa} were downloaded from \url{https://casa.nrao.edu/casa_obtaining.shtml}.}. There was no concern regarding visibility weights as all publicly available pipeline releases were after \gls{casa} 4.2.2 which contained a major correction to the handling of the weights.

Once calibrated data were obtained we inspected the visibilities from the calibrator sources to ensure the results of the calibration were as expected and all seriously problematic data were flagged. In most cases this inspection did not reveal anything that needed to be done beyond the observatory-provided calibration process. A few cases, however, did expose situations where marginal antennas or edge channels should have been flagged and we did so before imaging.
\subsection{Imaging} \label{imaging}
All imaging steps, from inspecting the visibilities up to and including cleaning the maps, were carried out in \gls{casa} 4.7.2\footnote{Note that after this paper was submitted, \gls{naasc} memo \num{117} was released detailing a problem with the flux scaling in \SI{7}{\metre} \gls{alma} mosaics imaged in versions of the \gls{casa} software prior to 5.0. We have measured the differences in clump fluxes between the \gls{dor} maps presented here and ones produced with \gls{casa} version 5.4.0-68 and find that on average, the Band 6 flux densities are overestimated by \num{\sim 8} per cent and Band 3 by \num{\sim 2} per cent. These small corrections have not been applied to the data presented in this paper.} (r39762) for all fields. Strong emission lines were identified with the plotms task and flagged before the visiblities were imaged to produce continuum maps. Since we combined maps in the two bands to create a dust-only emission map, we needed to match the spatial scales to which each pair of maps is sensitive as closely as possible. Matching the spatial scale sensitivities was achieved through a combination of trimming the shortest uv-spacings and tapering the weighting of the longest uv-spacings. To choose the ``inner'' uv trimming to use (the minimum uv-distance to include in imaging), we plotted the source visibility amplitudes versus their radial uv-distance in wavelengths for each band. The larger minimum baseline length of the two was used as the minimum uv-distance to include. For our fields, this minimum uv-distance was always set by the Band 6 data, so the uv-distance trimming was always applied to the Band 3 data.

The ``outer'' uv tapering directly changes the synthesized beam and sets the smallest spatial scale in the cleaned maps. The Band 6 maps always had smaller initial synthesized beams than the Band 3 maps. Once the uv-distance trimming was applied to the Band 3 data, we used the Band 3 synthesized beam as the target beam shape for the Band 6 uv  tapering.

Matching the beam shapes precisely through tapering alone was not possible given the differences in intrinsic uv coverage between the observations. Therefore, we used the \texttt{restoringbeam} parameter in \texttt{clean} to force the Band 6 Gaussian beam shape to be the same as that of the Band 3 map. Note that this means all noise and any emission left in the residual maps is still at the tapered resolution and only the emission that was cleaned is exactly matched to the Band 3 resolution. However, we were always able to bring the beams into fairly close agreement with the uv-tapering. Differences between the Band 6 and Band 3 beam axes were \num{<10} per cent (and usually much less) and position angle differences were \num{<5} per cent. Only \gls{dor} had a larger position angle difference (\SI{{\sim}30}{\degree} versus \SI{{\sim}60}{\degree}).

We cleaned all the dirty maps that showed obvious emission, which meant N166, GMC 225, and PCC~11546 were not cleaned. Motivated by the complexity of the emission in the \gls{dor} maps and by the desire to have a reproducible method of producing cleaned maps, we implemented an auto-masking algorithm for all cleaning. This algorithm is heavily based upon the auto-masking code given in the M100 \gls{casa} Guide as it appeared in August, 2016. We modified it to work as an automated \gls{casa} script\footnote{A full code listing of the auto-masking algorithm is available in Appendix B of Nathan Brunetti's MSc thesis at \url{https://macsphere.mcmaster.ca/handle/11375/22656}.}: when given a set of \texttt{clean} parameters, a minimum threshold to clean down to, a \acrlong{crtf} specifying an emission-free region of the dirty map, and a minimum spatial size to mask, the script iteratively cleaned down to the desired threshold. All auto-masking was done with stopping thresholds between \num{1.5} and \num{3} times the noise in the map and using minimum mask areas of \num{0.5} times the map synthesized beam. All fields were cleaned with Briggs weighting \citep{Bri1995}\footnote{Also see Dan Briggs' PhD thesis at \url{http://www.aoc.nrao.edu/dissertations/dbriggs/} for the full details of the weighting scheme and for more information on the \texttt{robust} parameter.}, with the \texttt{robust} parameter set to \num{0.5}, in multi-frequency synthesis mode, with \texttt{psfmode} set to ``psfclark'' and using a maximum of \num{e4} iterations. Table~\ref{tab:clean_summary} lists the values used for important \texttt{clean} parameters for each field. The average synthesized beam size was \SI{{\sim}2.3}{\arcsecond} corresponding to \SI{{\sim}0.6}{\parsec} at the distance to the \gls{lmc}. Mapped areas cover between \num{44} (N113) and \SI{1600}{\square\parsec}.

\begin{table*}
\begin{threeparttable}
\caption{Important \texttt{clean} parameters used in imaging.}
\label{tab:clean_summary}
\sisetup{table-number-alignment=center}
\begin{tabular}{@{}c S[table-format=1.3] c S[table-format=1.2] S[table-format=4.0] S[table-format=1.2] S[table-format=1.1] S[table-format=1.2] S[table-format=1.2] S[table-format=2.0]@{}}
\hline
           &                          &                             &                               &                    &                  &                 & \multicolumn{3}{c}{{\texttt{outertaper}}\tnote{c}} \\
\cmidrule{8-10}
           & {\texttt{cell}\tnote{a}} & {\texttt{imsize}\tnote{a}}  & {\texttt{threshold}\tnote{a}} & {\texttt{uvrange}} & {\texttt{minpb}} & \texttt{robust} & {Major}                  & {Minor}                  & {Position} \\
Field Name & {(\si{\arcsecond})}      & {(pixels)}                  & {(mJy/bm)}                    & {(${>}\lambda$)}   &                  &                 & {Axis (\si{\arcsecond})} & {Axis (\si{\arcsecond})} & {Angle (\si{\degree})} \\
\hline
\gls{dor}  & 0.18                     & 1152, 750                   & {0.30, 0.78}                  & 5092               & 0.38             & 0.5             & 0.85                     & 0.1                      & 67 \\
N159W      & 0.213                    & 1000, 720                   & {0.96, 1.32}                  & 8023               & 0.4              & 0.5             & 2.76                     & 1.64                     & 81 \\
N159E      & 0.21                     & 1250, 750                   & {0.80, 1.78}                  & 5956               & 0.5              & 0.5             & 3.05                     & 1.45                     & -70 \\
N113       & 0.13                     & ''                          & 0.40                          & {\ldots}           & 0.5              & 0.5             & {\ldots}                 & {\ldots}                 & {\ldots} \\
GMC 225    & {0.32, 0.11}             & 700, 1500                   & {\ldots}                      & {\ldots}           & 0.5              & 2               & {\ldots}                 & {\ldots}                 & {\ldots} \\
N166       & {0.38, 0.155}            & 1050, 1568                  & {\ldots}                      & {\ldots}           & 0.5              & 2               & {\ldots}                 & {\ldots}                 & {\ldots} \\
PCC 11546  & 0.126                    & 1680 $\times$ 2400\tnote{b} & {\ldots}                      & {\ldots}           & 0.5              & 2               & {\ldots}                 & {\ldots}                 & {\ldots} \\
\hline
\end{tabular}
\begin{tablenotes}
\item [] \textsc{Note.}---Columns are named after \texttt{clean} task parameter names.
\item [a] For fields with two entries the Band 3 value is reported first and Band 6 second.
\item [b] Square maps were made for all fields except PCC~11546.
\item [c] Applied to the Band 6 data only.
\end{tablenotes}
\end{threeparttable}
\end{table*}

\begin{table*}
\begin{threeparttable}
\caption{Image properties for each \gls{lmc} field.}
\label{tab:field_summary3}
\sisetup{table-number-alignment=center}
\begin{tabular}{@{}c S[table-format=1.2] S[table-format=1.2] S[table-format=1.1] S[table-format=1.2] S[table-format=1.2] S[table-format=2.0] S[table-format=1.2] S[table-format=1.2] S[table-format=1.2]@{}}
\hline
           & \multicolumn{2}{c}{{Map Area}\tnote{a}}           &                        & \multicolumn{3}{c}{{Synthesized Beam}\tnote{b}}              & \multicolumn{3}{c}{{Noise\tnote{d}}} \\
\cmidrule{2-3}\cmidrule(l{4pt}r{2pt}){5-7}\cmidrule(l{2pt}r{2pt}){8-10}
           & {Band 3}                & {Band 6}                & {Bandwidth\tnote{b,c}} & {Major}             & {Minor}             & {P. A.}          & {Dust}   & {Band 3}   & {Band 6} \\
Field Name & {(Sq. \si{\arcminute})} & {(Sq. \si{\arcminute})} & {(\si{\giga\hertz})}   & {(\si{\arcsecond})} & {(\si{\arcsecond})} & {(\si{\degree})} &          & {(mJy/bm)} \\
\hline
\gls{dor}  & 1.9                     & 1.2                     & 4                      & 2.25                & 1.40                & 65               & 0.41     & 0.15       & 0.39 \\
N159W      & 1.9                     & 1.6                     & {1.9, 1.6}             & 2.63                & 1.67                & 82               & 0.50     & 0.64       & 0.53 \\
N159E      & 1.8                     & 1.8                     & {1.9, 1.6}             & 2.90                & 1.61                & -77              & 0.66     & 0.32       & 0.60 \\
N113       & {\ldots}                & 0.21                    & 2.5                    & 1.35                & 1.04                & 61               & {\ldots} & {\ldots}   & 0.26 \\
GMC 225    & 4.0                     & 3.4                     & 1.9                    & {4.26, 2.01}        & {3.13, 1.13}        & {-67, -86}       & {\ldots} & 0.18       & 0.76 \\
N166       & 7.5                     & 4.0                     & 1.9                    & {4.08, 2.17}        & {3.57, 1.61}        & {-73, 69}        & {\ldots} & 0.20       & 0.94 \\
PCC 11546  & {\dots}                 & 5.8                     & 3.4                    & 1.81                & 1.23                & 75               & {\ldots} & {\ldots}   & 0.19 \\
\hline
\end{tabular}
\begin{tablenotes}
\item [] \textsc{Note.}---Synthesized beam shape parameters are given for the final images used to create our clump mass function.
\item [a] Area of the mapped field above a gain response threshold of 0.5. Dust-only maps are only defined where both Band 3 and Band 6 data overlap so dust-only map areas are the same as the Band 6 maps. At the distance of the \gls{lmc}, 1 square arcmin $\approx$ \SI{212}{\square\parsec}.
\item [b] For fields with two entries the Band 3 value is reported first and Band 6 second.
\item [c] Approximate bandwidth of continuum data used to produce the final maps. Bright emission lines are excluded in this estimate and in continuum imaging.
\item [d] Noise measurements were made in emission-free regions of the dirty map for each field.
\end{tablenotes}
\end{threeparttable}
\end{table*}

\subsection{Fields without detections}
We tried to improve the \gls{s/n} in the maps with no continuum emission by setting \texttt{robust} to \num{2.0} to favor \gls{s/n} over resolution. While the noise did drop, it was not enough to detect emission in any of the non-detection maps. The \acrshort{rms} noise and beam shapes listed in Table~\ref{tab:field_summary3} for GMC 225, N166, and PCC~11546 refer to the maps made with \texttt{robust=2.0}.

Almost half of the fields we investigated were not detected in continuum emission. This raises the question of whether those data are inherently of poorer sensitivity compared to the data that has abundant emission. We tested this by matching the synthesized beams across all fields to the largest beam in each band. At \SI{3.2}{\milli\metre} GMC 225 and N166 have the lowest \gls{rms} noise from our sample by a factor of 2-6. The \SI{1.3}{\milli\metre} observations of PCC~11546 have a smaller \gls{rms} noise than the fields with detections and the sensitivities of GMC 225 and N166 at this wavelength are only a factor of 2-3 worse. This comparison suggests there are intrinsic differences in the physical densities and structures of the non-detection fields compared to the four detected regions.

\section{Data analysis} \label{analysis}
\subsection{Free-free correction} \label{f-f_correction}
Given the frequencies of these observations we expect the main contributions to continuum emission to be thermal emission from the $\sim$\SI{30}{\kelvin} dust and free-free emission from ionized gas.
Although synchrotron emission is another possible source of continuum emission at these frequencies, HII regions, protostars, and young stars do not produce much synchrotron emission \citep{Gin2016}. Colliding-wind binaries have been observed with spectral indices indicative of synchrotron radiation from particles accelerating in the wind collision zone \citep{DeB2013}. However, following the method of \citet{Gin2016}, we estimate a \SI{3.2}{\milli\metre} flux density of \num{{\sim}6}~$\mu$Jy per binary. These flux densities are \SIrange{30}{100}{} times smaller than the noise in our fields (Table~\ref{tab:field_summary3}).

\begin{figure}
\centering
\includegraphics{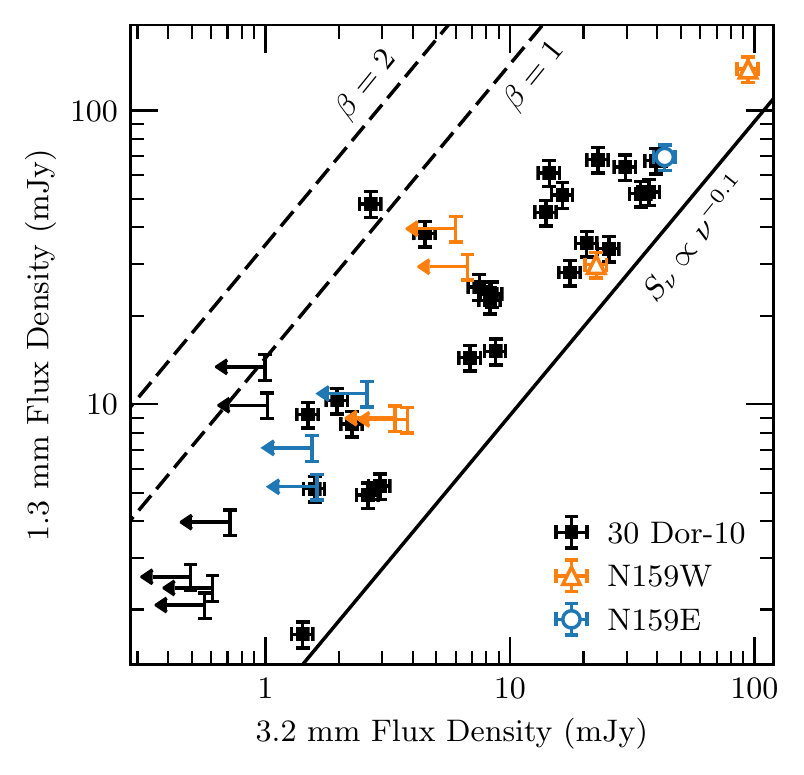}
\caption{Comparison of \SI{3.2}{\milli\metre} (Band 3) to \SI{1.3}{\milli\metre} (Band 6) flux densities for clumps identified at \SI{1.3}{\milli\metre}. Dashed lines show two different thermal dust emission scaling relations covering the range of realistic dust emissivity spectral indices, and the solid line shows a typical free-free scaling relation. Upper limits at \SI{3.2}{\milli\metre} are for clumps found at \SI{1.3}{\milli\metre} without significant co-spatial emission at \SI{3.2}{\milli\metre}. Error bars correspond to \num{10} per cent calibration uncertainty.}
\label{fig:b3_vs_b6}
\end{figure}

Fig.~\ref{fig:b3_vs_b6} compares the \num{1.3} and \SI{3.2}{\milli\metre} fluxes for clumps identified in the \SI{1.3}{\milli\metre} map. The solid line shows the flux scaling for free-free emission with exponent \num{-0.1} and two dashed lines show dust emission with exponents $(2+\beta)$ for $\beta=1$ and $\beta=2$. Since the points are clustered between the free-free and dust scaling relations we conclude that there is significant free-free emission even at \SI{1.3}{\milli\metre}. To use the dust emission to estimate the total mass of the clumps in these fields we need to correct for contamination from free-free emission at \SI{1.3}{\milli\metre}.

\begin{figure*}
\centering
\subfloat[\gls{dor}]{\includegraphics{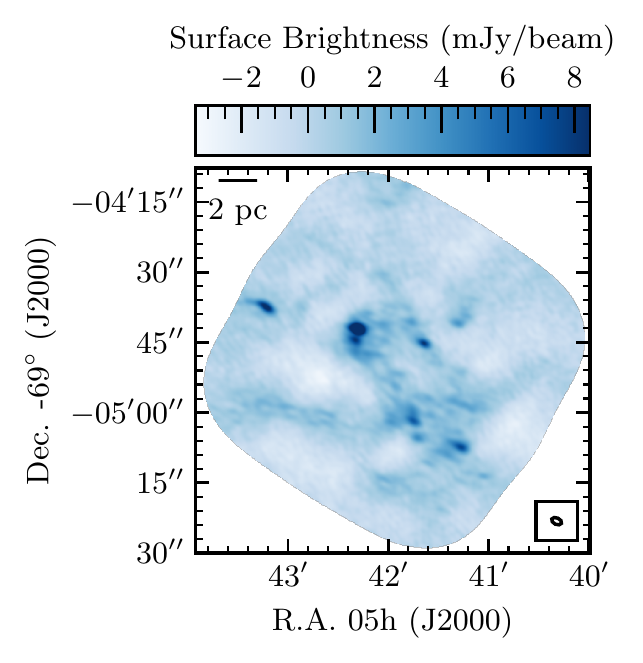}}
\subfloat[N159W]{\includegraphics{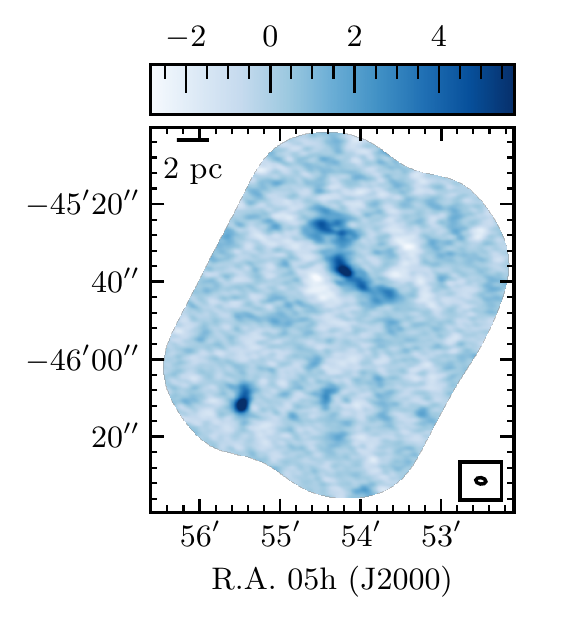}}
\subfloat[N159E]{\includegraphics{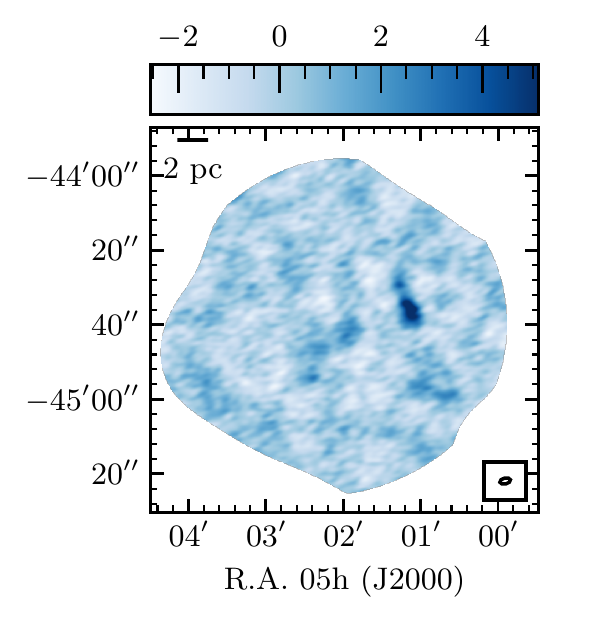}}
\caption{Dust-only maps of \emph{(a)} \gls{dor}, \emph{(b)} N159W, and \emph{(c)} N159E made from the combination of \SI{3.2}{\milli\metre} and \SI{1.3}{\milli\metre} \gls{alma} observations. Contamination from free-free emission has been removed from these maps using Equation~\ref{eq:b6_dust_only}.}
\label{fig:dust_maps}
\end{figure*}

If we assume that the dominant contributions are dust and free-free emission we can write
\begin{equation}
\label{eq:b6_just_b6}
S_{\mathrm{B6}} = S_{\mathrm{d}} + S_{\mathrm{ff}}
\end{equation}
where $S_{\mathrm{B6}}$ is the Band 6 flux density, $S_{\mathrm{d}}$ is the dust-only flux density, and $S_{\mathrm{ff}}$ is the free-free-only flux density, all measured at \SI{1.3}{\milli\metre}. For the Band 3 emission we can scale the dust and free-free emission from the Band 6 emission to obtain
\begin{equation}
\label{eq:scaled_b3}
S_{\mathrm{B3}} = S_{\mathrm{d}}\left(\frac{\nu_{\mathrm{B3}}}{\nu_{\mathrm{B6}}}\right)^{(2+\beta)} + S_{\mathrm{ff}}\left(\frac{\nu_{\mathrm{B3}}}{\nu_{\mathrm{B6}}}\right)^{-0.1}
\end{equation}
where $S_{\mathrm{B3}}$ is the Band 3 flux density, $\nu_{\mathrm{B3}}$ is the central frequency in Band 3, $\nu_{\mathrm{B6}}$ is the central frequency in Band 6, and $\beta$ is the emissivity spectral index of the dust. While we expect the Band 6 and Band 3 emission to come mostly from dust and free-free emission sources, respectively, including terms for both in Equations~{\ref{eq:b6_just_b6}} and~{\ref{eq:scaled_b3}} maintains the flexibility for the data to determine the relative fractions at each wavelength. The expression for the dust-only emission at Band 6 from solving Equations~\ref{eq:b6_just_b6} and~\ref{eq:scaled_b3} is then
\begin{equation}
S_{\mathrm{d}} = S_{\mathrm{B6}} - \frac{S_{\mathrm{B3}} - S_{\mathrm{B6}}\left(\frac{\nu_{\mathrm{B3}}}{\nu_{\mathrm{B6}}}\right)^{(2+\beta)}}{\left(\frac{\nu_{\mathrm{B3}}}{\nu_{\mathrm{B6}}}\right)^{-0.1} - \left(\frac{\nu_{\mathrm{B3}}}{\nu_{\mathrm{B6}}}\right)^{(2+\beta)}}.
\label{eq:b6_dust_only}
\end{equation}
We combine the Band 3 and Band 6 images using Equation~\ref{eq:b6_dust_only} to produce dust-only emission maps for our fields. We use an average value for the dust emissivity spectral index of $\beta=1.49$ from the \citet{Gor2014} \gls{lmc}-wide dust \gls{sed} fits to Herschel observations. Specifically we use the \gls{bembb} fit with unconstrained spectral index from \citet{Gor2014}. Dust-only maps for \gls{dor}, N159W, and N159E are shown in Fig.~\ref{fig:dust_maps}.

We tested our algebraic decomposition using lower frequency radio observations that are often used to measure free-free and synchrotron emission directly. \SI{4.8}{\giga\hertz} (\SI{6}{\centi\metre}) and \SI{8.4}{\giga\hertz} (\SI{3}{\centi\metre}) observations were made of 30~Doradus \citep{Laz2003} and N159 \citep{Ind2004} with the \gls{atca}. We applied spatial filtering to the radio images, to match the \gls{aca} uv coverage of the \gls{alma} observations, through the \gls{casa} observation simulation functionality. This allowed us to better match the spatial scales measured at the different frequencies (especially for \gls{dor} where the \gls{aca} uv coverage filtered out about 50 per cent of the radio flux).

We first measured total fluxes in each field at \SI{4.8}{\giga\hertz} and \SI{8.4}{\giga\hertz}. Adopting a synchrotron spectral index of \num{-0.83} and a free-free spectral index of \num{-0.1}, we iterated to estimate the balance of the two emission sources at the two frequencies. We then extrapolated the free-free component to the \gls{alma} Band 3 frequency to get an estimate of the expected free-free emission at \SI{3.2}{\milli\metre}.

To compare to this estimate, we solved for the free-free-only emission at Band 3 (see Appendix~\ref{all_maps} for specific decomposition expressions and maps of free-free-only emission). Assuming a \num{10} per cent absolute flux calibration uncertainty on the radio continuum, the total free-free emission in \gls{dor} matches within \num{15} per cent between extrapolated and decomposed estimates (\num{0.82}~Jy and \num{0.97}~Jy, respectively). N159W is consistent within the radio uncertainties, with \num{75}~mJy and \SI{78.5}~mJy in the ATCA and \gls{alma} maps, respectively.

N159E appears to have about four times too much free-free flux at Band 3 compared to the extrapolated flux (it has a near synchrotron-only spectral index in the radio maps). If the free-free emission becomes optically thick between \SI{8.6}{\giga\hertz} and \SI{98}{\giga\hertz} then the free-free spectrum would turn over \citep[e.g.][]{Tur1998} leaving just the synchrotron spectral index and a low free-free flux at low frequencies. Galactic ultracompact HII regions have been observed with free-free spectral turnovers at frequencies between \num{10} and \SI{20}{\giga\hertz} \citep{Woo1989}. It is possible that the main region of bright radio emission is originating from a population of ultracompact HII regions averaged together with supernova remnants within the \gls{alma} beam leading to synchrotron-dominated emission near \SI{10}{\giga\hertz} and the presence of free-free emission at and above \SI{100}{\giga\hertz}.

Lastly, we estimate that diffuse synchrotron emission contributes up to ${\sim}$\num{12} per cent of the Band 3 flux. Since our method scales between Band 3 and 6 with the free-free spectral index this synchrotron component would be oversubtracted from the Band 6 flux by Equation~\ref{eq:b6_dust_only}. This results in, at most, four per cent of the total Band 6 flux being oversubtracted when making the dust-only maps by neglecting a synchrotron term in Equation~\ref{eq:b6_dust_only}. Considering the imperfect uv coverage matching between the ATCA and \gls{alma} observations and the \num{10} per cent calibration uncertainty on the \gls{alma} and \gls{atca} observations, we believe the radio free-free emission estimates agree well enough and that the synchrotron is weak enough in Band 6 that we continue with our algebraic method.

We are not able to include N113 in our final mass function analysis because of the non-trivial free-free emission illustrated in Fig.~\ref{fig:b3_vs_b6} and our lack of observations at \SI{3.2}{\milli\metre} for this field. Not being able to include N113 was unfortunate since it contributed about the same number of clumps in Band 6 as the N159E map, even within the much smaller area.
\subsection{Clump finding} \label{clump_finding}
To identify molecular gas clumps in our dust-only maps we used the ClumpFind algorithm \citep{Wil1994} available in Starlink\footnote{The Starlink software \citep{Cur2014} is currently supported by the East Asian Observatory.} through the Cupid package \citep{Ber2007}. We chose to use this algorithm to facilitate direct comparison with previous work that also used ClumpFind. A total of \num{32} dust clumps were identified in three fields (\gls{dor}, N159W and N159E) with the majority found in \gls{dor}. Clump properties are summarized in Table~\ref{tab:clump_props}.

The \gls{rms} noise we input to ClumpFind was measured in each dust-only map from a region where no obvious emission, artifacts or negative bowls were visible. The lowest contour was set to three times the map \gls{rms} to ensure sources could be trusted with some confidence while also trying to recover most of the emission associated with a clump. Contour spacings were set to two times the \gls{rms} as this minimized splitting up sources while still recovering most of the obvious features. This combination of lowest contour level and contour spacing was also shown to result in the lowest number of missed and ``false'' clumps through synthetic source detection testing by \citet{Wil1994}. Our own testing with several choices for these two settings showed that our final choice gave results that picked out distinct but somewhat blended sources without extracting spurious looking sources, along with staying well within believable bounds around obvious sources. Background subtraction was not used because the interferometer's intrinsic spatial filtering removes large-scale emission, while the complicated interplay of bright emission can produce adjacent negative regions.

During completeness testing (\S~\ref{completeness}) we found that our setting for the beam \gls{fwhm} parameter in ClumpFind caused the algorithm to reject sources as too small when it should not have done so. To avoid this problem we instead set the \gls{fwhm} parameter to zero and specified the \texttt{minpix} parameter to reject clumps that were too small. To determine the correct \texttt{minpix} value, we ran ClumpFind once with the \gls{fwhm} set to the geometric mean of the synthesized beam axes and noted the \texttt{minpix} value ClumpFind calculated internally: \num{56} for \gls{dor} and N159W and \num{61} for N159E. This method gave the same resulting clumps as when using the \gls{fwhm} parameter but it prevented clumps near the size of the beam from being rejected as too small due to the lowest contour setting (Berry, private communication). As this could select clumps that were of the same area as the synthesized beam but smaller than one of the beam dimensions (e.g. long and skinny clumps) we inspected the clump regions identified with ClumpFind visually for each field.

\begin{table*}
\begin{threeparttable}
\caption{Properties of clumps in the \gls{lmc}.}
\label{tab:clump_props}
\sisetup{table-number-alignment=center,separate-uncertainty=true,table-sign-mantissa}
\begin{tabular}{@{}c c c S[table-format=1.2] S[table-format=2.1] S[table-format=2.1(1)] S[table-format=2.1] S[table-format=2.2(2)] S[table-format=4.0(3)]@{}}
\hline
             & \acrshort{ra}\tnote{a} & \acrshort{dec}\tnote{a} & {Area\tnote{b}}         & {$S_{\mathrm{ff,peak}}$\tnote{c,d}} & {$S_{\mathrm{ff,int}}$\tnote{c,e}} & {$S_{\mathrm{d,peak}}$\tnote{f,g}} & {$S_{\mathrm{d,int}}$\tnote{f,h}} & {$M$\tnote{i}} \\
Name         & (J2000)                & (J2000)                 & {(\si{\square\parsec})} & {(mJy/beam)}                    & {(mJy)}                        & {(mJy/beam)}                   & {(mJy)}                        & {(\si{\solarmass})} \\
\hline
\gls{dor} 1  & 05:38:49.22            & -69:04:42.24            & 2.6                     & 3.7                             & 21.5 \pm 0.5                   & 19.7                           &  54 \pm 2                      & 5560(160) \\
\gls{dor} 2  & 05:38:52.84            & -69:04:37.55            & 1.8                     & 0.4                             & 0.6 \pm 0.4                    & 9.5                            &  27 \pm 1                      & 2750(140) \\
\gls{dor} 3  & 05:38:46.56            & -69:04:45.30            & 1.5                     & 0.9                             & -0.9 \pm 0.4                   & 8.3                            &  18 \pm 1                      & 1890(120) \\
\gls{dor} 4  & 05:38:49.32            & -69:04:44.40            & 3.4                     & 1.9                             & 6.2 \pm 0.6                    & 8.3                            &  45 \pm 2                      & 4650(180) \\
\gls{dor} 5  & 05:38:45.08            & -69:05:07.26            & 3.7                     & 4.0                             & 25.2 \pm 0.6                   & 7.5                            &  44 \pm 2                      & 4530(190) \\
\gls{dor} 6  & 05:38:47.00            & -69:05:01.68            & 4.9                     & 1.4                             & 5.4 \pm 0.7                    & 6.7                            &  56 \pm 2                      & 5740(220) \\
\gls{dor} 7  & 05:38:46.83            & -69:05:05.28            & 0.97                    & 0.8                             & 0.9 \pm 0.3                    & 4.2                            &  9.8 \pm 0.9                   & 1012(96) \\
\gls{dor} 8  & 05:38:45.22            & -69:04:40.98            & 1.2                     & 0.6                             & -0.2 \pm 0.3                   & 3.8                            &  11 \pm 1                      & 1110(110) \\
\gls{dor} 9  & 05:38:47.06            & -69:04:40.80            & 1.4                     & 2.7                             & 6.6 \pm 0.4                    & 3.4                            &  12 \pm 1                      & 1230(120) \\
\gls{dor} 10 & 05:38:48.21            & -69.04.41.16            & 1.5                     & 4.8                             & 18.5 \pm 0.4                   & 3.1                            &  13 \pm 1                      & 1330(120) \\
\gls{dor} 11 & 05:38:44.98            & -69.04.58.26            & 3.1                     & 1.0                             & 7.1 \pm 0.6                    & 3.1                            &  27 \pm 2                      & 2760(170) \\
\gls{dor} 12 & 05:38:47.70            & -69.04.54.30            & 0.75                    & -0.3                            & -1.7 \pm 0.3                   & 2.7                            &  6.3 \pm 0.8                   & 654(85) \\
\gls{dor} 13 & 05:38:46.39            & -69.04.57.00            & 0.78                    & 0.5                             & 0.3 \pm 0.3                    & 2.6                            &  6.8 \pm 0.8                   & 705(87) \\
\gls{dor} 14 & 05:38:44.21            & -69.05.13.38            & 0.53                    & 0.9                             & 1.3 \pm 0.2                    & 2.6                            &  4.1 \pm 0.7                   & 424(72) \\
\gls{dor} 15 & 05:38:52.14            & -69.04.58.62            & 1.4                     & 1.8                             & 7.4 \pm 0.4                    & 2.5                            &  10 \pm 1                      & 1070(120) \\
\gls{dor} 16 & 05:38:48.14            & -69.05.14.10            & 1.5                     & 1.2                             & 2.7 \pm 0.4                    & 2.3                            &  11 \pm 1                      & 1150(120) \\
\gls{dor} 17 & 05:38:47.60            & -69.04.51.78            & 0.53                    & 1.3                             & 0.7 \pm 0.2                    & 2.3                            &  4.3 \pm 0.7                   & 447(71) \\
\gls{dor} 18 & 05:38:50.29            & -69.05.00.42            & 1.4                     & 1.6                             & 8.2 \pm 0.4                    & 2.2                            &  10 \pm 1                      & 1060(120) \\
\gls{dor} 19 & 05:38:45.52            & -69.04.59.70            & 0.64                    & 1.0                             & 1.3 \pm 0.3                    & 2.2                            &  5.1 \pm 0.8                   & 524(78) \\
\gls{dor} 20 & 05:38:53.05            & -69.04.57.53            & 1.6                     & 2.9                             & 12.4 \pm 0.4                   & 2.2                            &  12 \pm 1                      & 1220(130) \\
\gls{dor} 21 & 05:38:46.36            & -69.05.10.68            & 0.31                    & 1.5                             & 1.2 \pm 0.2                    & 2.1                            &  2.4 \pm 0.5                   & 249(55) \\
\gls{dor} 22 & 05:38:45.96            & -69.04.58.44            & 0.24                    & 0.2                             & 0.0 \pm 0.2                    & 2.1                            &  2.0 \pm 0.5                   & 205(48) \\
N159W 1      & 05:39:41.90            & -69.46.11.48            & 2.0                     & 0.8                             & -1 \pm 2                       & 8.4                            &  25 \pm 2                      & 2610(230) \\
N159W 2      & 05:39:36.77            & -69.45.37.41            & 2.7                     & 1.5                             & 2 \pm 2                        & 7.6                            &  31 \pm 3                      & 3200(260) \\
N159W 3      & 05:39:37.96            & -69.45.25.27            & 3.3                     & 21.8                            & 54 \pm 2                       & 4.6                            &  28 \pm 3                      & 2890(290) \\
N159W 4      & 05:39:35.95            & -69.45.41.03            & 1.4                     & 0.1                             & -1 \pm 1                       & 4.2                            &  12 \pm 2                      & 1270(190) \\
N159W 5      & 05:39:36.93            & -69.45.27.40            & 1.5                     & 18.7                            & 15 \pm 2                       & 4.0                            &  13 \pm 2                      & 1290(200) \\
N159W 6      & 05:39:34.63            & -69.45.42.94            & 1.5                     & 0.3                             & -1 \pm 1                       & 3.3                            &  11 \pm 2                      & 1150(190) \\
N159W 7      & 05:39:37.75            & -69.46.09.57            & 0.73                    & 5.5                             & 6 \pm 1                        & 2.8                            &  5 \pm 1                       & 530(140) \\
N159E 1      & 05:40:04.47            & -69.44.35.88            & 1.8                     & 11.3                            & 26.9 \pm 0.7                   & 6.6                            &  22 \pm 2                      & 2320(180) \\
N159E 2      & 05:40:04.71            & -69.44.34.41            & 0.75                    & 2.2                             & 0.7 \pm 0.5                    & 6.0                            &  9 \pm 1                       & 900(120) \\
N159E 3      & 05:40:05.08            & -69.44.29.37            & 0.65                    & 0.0                             & -0.4 \pm 0.5                   & 3.9                            &  6 \pm 1                       & 570(110) \\
\hline
\end{tabular}
\begin{tablenotes}
\item [a] Positions are peak positions as reported by ClumpFind.
\item [b] Calculated from the ``square arcseconds'' value reported by ClumpFind.
\item [c] Measured at \SI{1.3}{\milli\metre} after correcting for dust contamination using Equation~\ref{eq:b6_ff_only} to isolate for the free-free component.
\item [d] Uncertainties are random statistical uncertainties measured from each free-free-only map and are the same for all clumps in a given field: \num{0.2}~mJy/beam for \gls{dor}, \num{0.3}~mJy/beam for N159W and \num{0.3}~mJy/beam for N159E.
\item [e] Uncertainties are calculated by propagating uncertainties in each band through Equation~\ref{eq:b6_ff_only}. Uncertainties in each band are the \gls{rms} noise in each band times the square root of the number of beams covering the clump area in the free-free-only map.
\item [f] Measured at \SI{1.3}{\milli\metre} after correcting for free-free contamination using Equation~\ref{eq:b6_dust_only}.
\item [g] Uncertainties are random statistical uncertainties measured from each dust-only map and are the same for all clumps in a given field: \num{0.4}~mJy/beam for \gls{dor}, \num{0.8}~mJy/beam for N159W and \num{0.7}~mJy/beam for N159E.
\item [h] Uncertainties are calculated by propagating uncertainties in each band through Equation~\ref{eq:b6_dust_only}. Uncertainties in each band are the \gls{rms} noise in each band times the square root of the number of beams covering the clump area in the dust-only map.
\item [i] Calculated using Equation~\ref{eq:flux2mass} with $T_{\mathrm{d}}=\SI{36.7}{\kelvin}$, $\kappa_{\SI{1.3}{\milli\metre}}=\SI{1.16}{\square\centi\metre\per\gram}$, and \acrshort{g2d}$=\num{500}$.
\end{tablenotes}
\end{threeparttable}
\end{table*}

\subsection{Converting dust flux to total gas mass} \label{flux2mass}
We calculated total mass (gas plus dust) for each of the clumps identified in our dust-only maps. Clump flux densities were converted to dust masses using
\begin{equation}
\label{eq:flux2mass}
M_{\mathrm{d}} = \frac{S d^{2}}{\kappa B(T_{\mathrm{d}})}
\end{equation}
where $S$ is the dust flux density integrated over the spatial extent of the clump, $d$ is the distance to the \gls{lmc}, $\kappa$ is the dust opacity per unit mass column density, and $B(T_{\mathrm{d}})$ is the Planck function evaluated at the dust temperature $T_{\mathrm{d}}$. Given the distance measurement from \citet{Pie2013} of \SI{49.97 \pm 1.11}{\kilo\parsec} we adopted a distance of \SI{50}{\kilo\parsec} for our calculations. The dust temperature and opacity were calculated from the \citet{Gor2014} temperature and emissivity maps where we averaged the pixel values covering \gls{dor}, N159W, N159E and N113 to obtain a temperature of \SI{36.7}{\kelvin} and an opacity of \SI{1.16}{\square\centi\metre\per\gram}.

With the dust mass calculated using Equation~\ref{eq:flux2mass} we applied a conversion to total gas mass through a \gls{g2d} ratio of \num{500} from \citet{Rom2017}. This \gls{g2d} ratio was obtained by stacking the dust \glspl{sed} across the entire \gls{lmc} as measured with IRAS and Planck in bins of varying gas surface density. \citet{Rom2017} derived atomic gas surface densities from \SI{21}{\centi\metre} Parkes observations \citep{Sta2003} and molecular gas surface densities from $^{12}$\gls{co} (1-0) NANTEN observations \citep{Miz2001}. The stacked \glspl{sed} were fit with a modified blackbody to estimate the dust surface density, as well as the temperature and spectral emissivity index. We adopted the \gls{g2d} from the highest gas surface density bin from \citet{Rom2017} as this is likely where stars are forming. Applying this conversion to our dust \gls{rms} noise measurements from Table~\ref{tab:field_summary3} results in mass sensitivities (\num{1}$\sigma$) of \num{45}, \num{85}, and \SI{72}{\solarmass} for \gls{dor}, N159W, and N159E, respectively.

\begin{figure}
\centering
\includegraphics{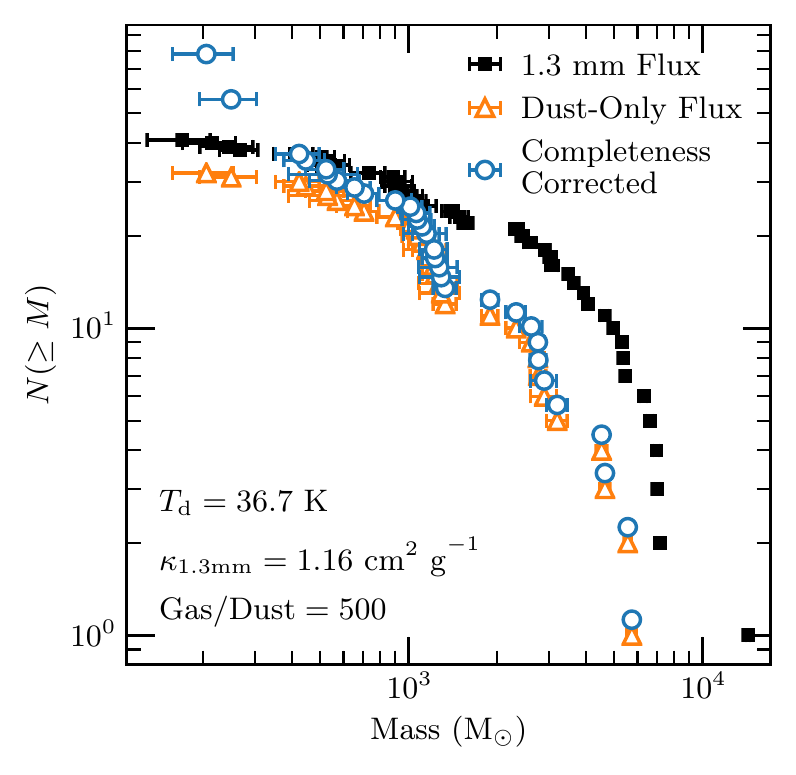}
\caption{Comparison of cumulative clump mass functions from \SI{1.3}{\milli\metre} maps (black squares), observed dust-only maps (orange triangles), and completeness corrected dust-only maps (blue circles). The parameters used to convert flux to gas mass in Equation~\ref{eq:flux2mass} are shown in the bottom left. Error bars are random uncertainties on clump masses from noise in the maps; see \S~\ref{flux2mass} for details.}
\label{fig:dust_and_b6_cmf}
\end{figure}

Fig.~\ref{fig:dust_and_b6_cmf} shows the cumulative dust-only clump mass function along with the mass function derived from the \SI{1.3}{\milli\metre} map assuming the \SI{1.3}{\milli\metre} emission is purely from dust. Fig.~\ref{fig:dust_and_b6_cmf} shows that our free-free correction is significantly larger than the random noise in each clump and systematically shifts the clump mass function to smaller masses.

\subsection{Completeness corrections} \label{completeness}
\begin{figure*}
\centering
\includegraphics{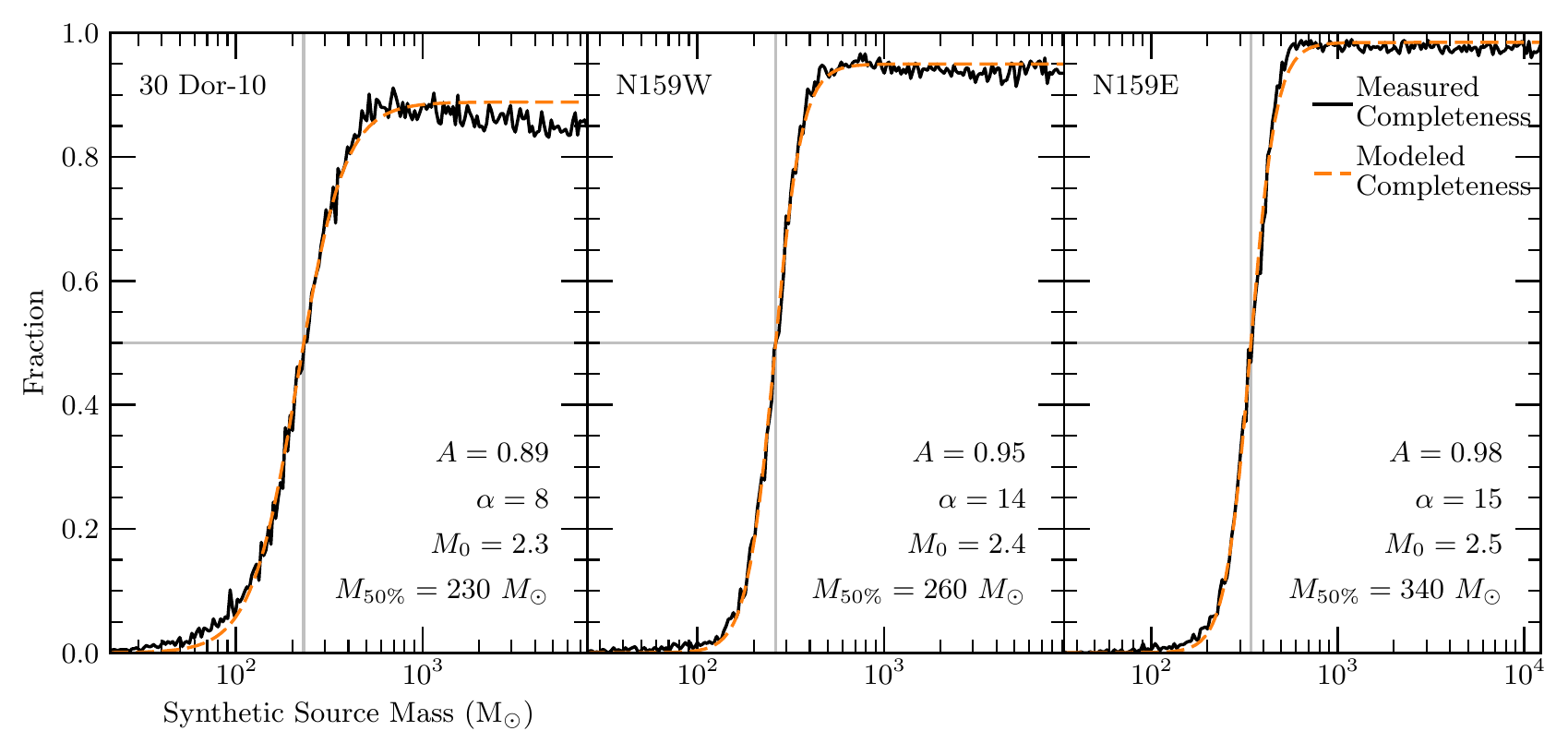}
\caption{Completeness curves for point sources in \gls{dor}, N159W, and N159E (black line). The best fit logistic curves (Equation~\ref{eq:logistic}) are shown as orange dashed lines with the best fit parameters in the bottom right. \num{50} per cent completeness is marked by the vertical gray lines and reported as $M_{50\%}$ in the bottom right.}
\label{fig:all_completeness_fit}
\end{figure*}

We estimated the completeness of our mass function through Monte Carlo techniques by injecting synthetic sources of known fluxes and sizes into the dust-only maps one source at a time. We ran ClumpFind on the altered map with the same parameters as for the original maps and counted when synthetic clumps were recovered and when they were not. We then took the ratio of the number recovered to the total number injected as the completeness fraction. Synthetic source positions were randomly placed across the maps with a uniform distribution. All sources were elliptical Gaussians with \gls{fwhm} and position angle identical to the synthesized beam for the corresponding map. We inject \num{600} synthetic sources per mass bin. Synthetic source masses were spaced evenly in log-space from roughly \SIrange{21}{1.2e4}{\solarmass}. Fig.~\ref{fig:all_completeness_fit} shows completeness data for \gls{dor}, N159W and N159E.

We fit each completeness curve with a standard logistic function \citep{Har2016,Fre2017}
\begin{equation}
f(M) = A\left\{1 + \exp{\left[-\alpha\left(\log_{10}\frac{M}{\mathrm{M}_{\odot}} - M_{0}\right)\right]}\right\}^{-1}
\label{eq:logistic}
\end{equation}
where $f(M)$ is the completeness fraction, $A$ is an amplitude to allow for curves that do not asymptote to one, $\alpha$ controls the width of the central part of the curve and $M_{0}$ is the logarithm of the mass at which $f = 0.5A$. Fig.~\ref{fig:all_completeness_fit} shows the best fit curves and parameters for each field's completeness data.

A surprising feature of the completeness curves shown in Fig.~\ref{fig:all_completeness_fit} is that none of the maps actually attain \num{100} per cent completeness. This is related to how crowded each map is: a synthetic source added to a crowded map may land on a real clump so that ClumpFind does not find the extra clump. \gls{dor} is the most crowded map and has the lowest maximum completeness, while N159E is the least crowded and has nearly \num{100} per cent recovery at high masses. This trend shows that not only the noise and resolution but also the crowdedness of our maps are limiting our ability to find clumps.

It is important to note that we only tested source recovery for sources that were the same size as the synthesized beam, while most of the clumps we have identified are much larger than the synthesized beam. This is a general challenge in studying these types of objects, as molecular clumps are intrinsically amorphous in shape and vary significantly in physical size. Since larger clumps of the same integrated flux density as smaller clumps have lower \gls{s/n}, our completeness estimates are likely overestimating the total fraction of clumps recovered for a given mass. While more sophisticated methods can be employed \citep[e.g. see Appendix B of][]{Kon2015} having a simplistic estimate of the completeness is better than no estimate.

We use the fits to the completeness data to correct the mass function numbers by calculating $N_{\mathrm{corr}} = N(\geq M)/f(M)$. We used the \gls{dor} completeness curve because it contributes the majority of the clumps in our mass function. Applying this completeness correction produces the mass function shown as blue points in Fig.~\ref{fig:dust_and_b6_cmf}, in comparison to the observed mass function shown in orange points.

\section{Mass functions in the LMC and Milky Way} \label{compare2MW}
\subsection{\gls{lmc} clump mass function} \label{LMC_CMF}
An empirical mass distribution is typically plotted as either a differential mass function or a cumulative mass function. A differential mass function is calculated by defining mass bins spanning the mass range of objects and counting the number within each bin. This form allows for a simple Poissonian counting statistics approach to uncertainties for fitting but the results can depend on the choices of bin widths and centers. A cumulative mass function avoids complications from binning as it is just an ordered tally of masses at or above a given object's mass. The trade off is that there is not a simple statistical approach to handling uncertainties for fitting.

A common model for the cumulative mass function of clumps and cores is a broken power law with parameters for two independent power law indices and a break mass. This model can be written as
\begin{equation}
N(\geq M) = 
\begin{cases}
AM_{\mathrm{break}}^{(\alpha_{\mathrm{high}} - \alpha_{\mathrm{low}})}M^{(\alpha_{\mathrm{low}}+1)} & M < M_{\mathrm{break}} \\
AM^{(\alpha_{\mathrm{high}}+1)} & M \geq M_{\mathrm{break}}
\end{cases}
\label{eq:dbl_pwr_law}
\end{equation}
where $A$ is an arbitrary amplitude, $M_{\mathrm{break}}$ is the break mass where the power law index changes, $\alpha_{\mathrm{low}}$ is the power law index for masses below $M_{\mathrm{break}}$, and $\alpha_{\mathrm{high}}$ is the power law index for masses above $M_{\mathrm{break}}$ \citep{Rei2006b}.

\begin{table*}
\begin{threeparttable}
\caption{Results of double power law and lognormal fits to cumulative clump mass function for the \gls{lmc}.}
\label{tab:clfit_summary}
\begin{tabular}{@{}l c c c c c@{}}
\hline
                               & $\alpha_{\mathrm{low}}$                  & $\alpha_{\mathrm{high}}$          & $M_{\mathrm{break}}$                  & $M_{0}$\tnote{a}                      & $\sigma$ \\
\hline
\multirow{2}{*}{Initial Guess} & \multirow{2}{*}{-1.8}                 & \multirow{2}{*}{-3.5}         & \multirow{2}{*}{2500}              & \multirow{2}{*}{7}                   & \multirow{2}{*}{1}\\ \\
Observed                       & \multirow{2}{*}{$-1.50^{+0.06}_{-0.05}$} & \multirow{2}{*}{$-3.3\pm0.2$} & \multirow{2}{*}{$2200\pm200$}      & \multirow{2}{*}{$7.22^{+0.06}_{-0.04}$} & \multirow{2}{*}{$0.86^{+0.03}_{-0.04}$} \\
Mass Function                  & \\
Observed MF,                   & \multirow{2}{*}{$-1.74^{+0.01}_{-0.1}$} & \multirow{2}{*}{$-3.3\pm0.2$}  & \multirow{2}{*}{$2500^{+300}_{-100}$} & \multirow{2}{*}{$7.3^{+0.2}_{-0.1}$}    & \multirow{2}{*}{$0.81^{+0.04}_{-0.06}$} \\
M > \SI{500}{\solarmass} Only  & \\
Completeness                   & \multirow{2}{*}{$-1.76^{+0.04}_{-0.05}$} & \multirow{2}{*}{$-3.3\pm0.2$} & \multirow{2}{*}{$2500^{+300}_{-200}$} & \multirow{2}{*}{$6.8\pm0.2$}          & \multirow{2}{*}{$1.03^{+0.06}_{-0.08}$} \\
Corrected MF                   & \\
\hline
\end{tabular}
\begin{tablenotes}
\item [a] $\exp{(M_{0})}$ is the central mass of the lognormal function from Equation~\ref{eq:log_norm}. So, for example, the fitted central mass of the observed mass function is $\exp{(7.22)} \approx$ \SI{1370}{\solarmass}.
\end{tablenotes}
\end{threeparttable}
\end{table*}

Given our relatively small sample of clumps and the simple model, we carried out our fits with the standard nonlinear Levenberg-Marquardt least-squares minimization \citep{Mor1977}. To estimate fitting weights we followed the approach used by \citet{Rei2006a} which specifies the y-data uncertainties as the cumulative number $N({\geq}M)$ for each clump. The best fit double power law parameters are given in Table~\ref{tab:clfit_summary} for the mass function determined directly from the clump masses along with two variations that attempt to account for incompleteness. The procedure and inputs for the fits are the same across all three variations.

Given the form of the fitting function, there are only uncertainties in the x-values. To estimate the absolute uncertainties on the best-fit parameters we used Monte Carlo simulation, as done by \citet{Rei2006a}. This involved generating \num{e5} artificial mass functions and fitting each in the same way as the observed mass function. The artificial mass functions were made by drawing normally distributed random deviates centered on each measured clump mass and with standard deviations equal to the measurement uncertainties on the masses from the dust-only maps. Each newly generated sample of \num{32} masses was then sorted into descending order to prepare for fitting. Not only does this Monte Carlo approach simplify estimating the fit parameter uncertainties but it also nicely includes the effects of neighboring masses swapping places in the sorted order due to their uncertainties. From the distributions of fit parameters we used the inner \num{68} per cent to obtain the 1$\sigma$ uncertainties on each fit parameter (Table~\ref{tab:clfit_summary}).

\begin{figure*}
\centering
\includegraphics{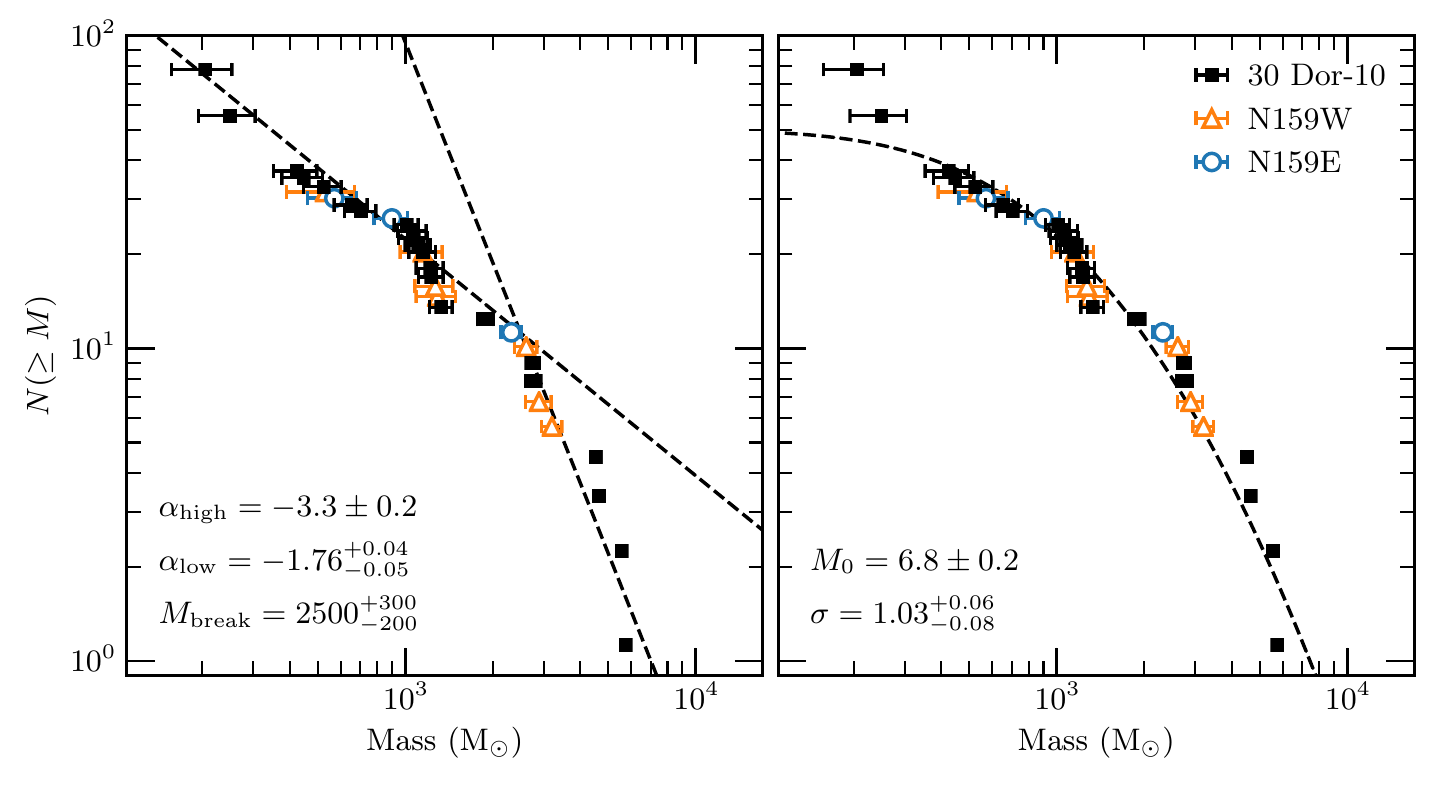}
\caption{Completeness corrected clump mass function (points are the same as the blue circles in Fig.~{\ref{fig:dust_and_b6_cmf}}) with best fits plotted as black dashed lines; double power law on left and lognormal on right. Parameters of the fits (Equations~\ref{eq:dbl_pwr_law} and \ref{eq:log_norm}) are shown in the bottom left. Clumps are color coded by the region from which they originate with \mbox{\gls{dor}} as black squares, N159W as orange triangles, and N159E as blue circles. Error bars are the same as in Fig.~{\ref{fig:dust_and_b6_cmf}}.}
\label{fig:corrected_cmf}
\end{figure*}

Fig.~\ref{fig:corrected_cmf} shows the completeness-corrected mass function with the best fit double power law (Table~\ref{tab:clfit_summary}). We use all clumps in our sample in this fitting with our lowest mass of \SI{205}{\solarmass} at the \num{41} per cent complete level. We also tested restricting the fit to masses above \SI{500}{\solarmass} in the uncorrected mass function since this is roughly where the \gls{dor} completeness curve flattens out at maximum completeness. The best fit double power laws for the two methods end up being statistically indistinguishable based on our Monte Carlo uncertainty estimates (Table~\ref{tab:clfit_summary}) so we focus our analysis on the completeness corrected mass function.

It may appear unusual that the numbers of high mass clumps need to be boosted by our completeness correction. An overlapping pair of sources (merged clumps) would be identified in the data as a more massive clump and thus not be entirely missed. We tested removing the completeness correction to the high mass clumps by stitching together two sigmoid curves (the fit shown in Fig.~\ref{fig:all_completeness_fit} and a curve with the same width parameter but amplitude of \num{1.0} and shifted slightly to higher masses) to create a smoothly varying sigmoid matching our measured completeness at low masses and asymptoting to \num{100} per cent complete at high masses. Refitting the mass function corrected with this piece-wise completeness curve did steepen both indices and moved the break to a higher mass but all within the uncertainties of our original fit.

For comparison with studies that fit mass distributions with a lognormal function we have also fit our completeness-corrected \gls{lmc} mass function with
\begin{equation}
N(\geq M) =\frac{A}{2}\left[ 1 - \mathrm{erf}\left(\frac{\ln M - M_{0}}{\sqrt{2}\sigma}\right)\right]
\label{eq:log_norm}
\end{equation}
where $A$ is an arbitrary amplitude, $\exp{(M_{0})}$ is the central mass, and $\sigma$ is a width parameter. Best-fit parameters are summarized in Table~\ref{tab:clfit_summary} along with 1$\sigma$ uncertainties. All fitting and uncertainty estimation was carried out identically to the double power law procedure.

\subsection{Comparison to other studies}
\begin{figure}
\centering
\includegraphics{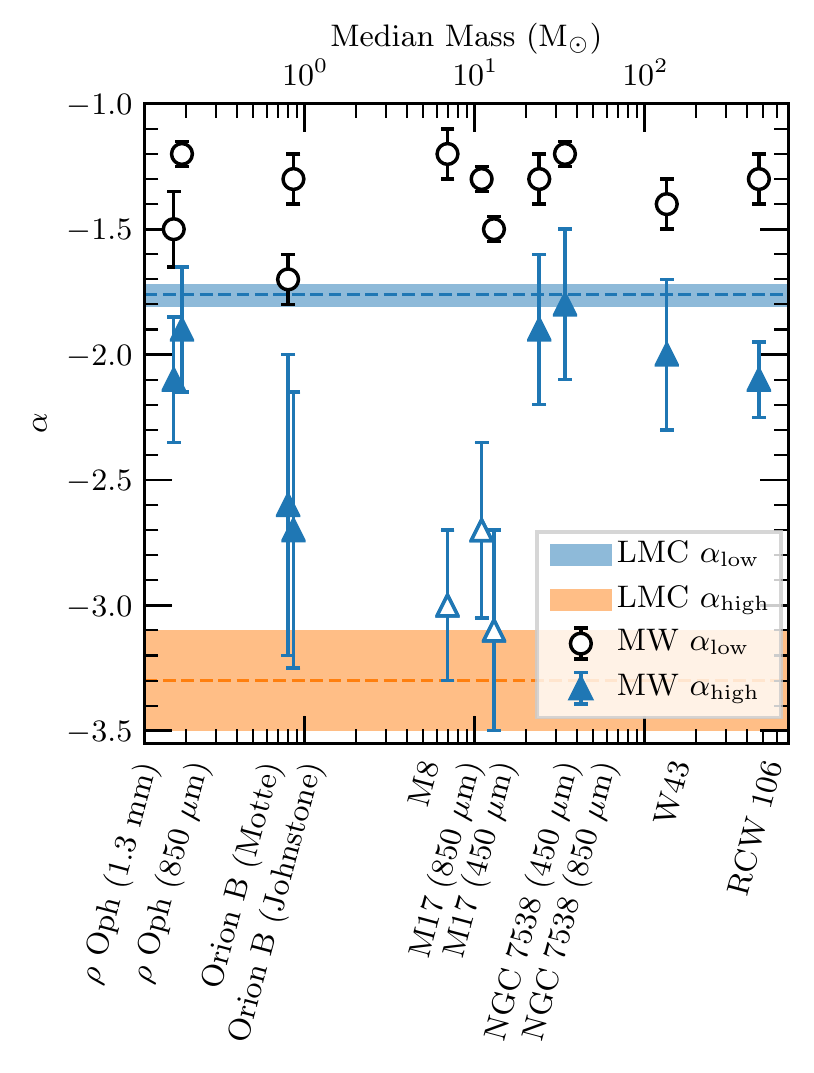}
\caption{Comparison of the double power law mass function fits from the seven Galactic star forming regions summarized in \citet{Rei2006b} with the low and high mass fit indices for our \gls{lmc} mass function. Circles and triangles show the Galactic low and high mass indices, respectively. The orange and blue dashed lines and shaded regions show the low and high mass indices for the \gls{lmc}, respectively. Milky Way clouds are arranged with their median core/clump masses increasing to the right. Open triangles denote less robust high mass slope determinations for Galactic clouds M8 (clumps identified by eye) and M17 (mass function fit better by lognormal function).}
\label{fig:alpha_compare}
\end{figure}

The sample of \num{11} core/clump mass functions from \num{7} different star forming regions analyzed by \citet{Rei2006b} is a useful one for comparing our \gls{lmc} clump mass function to Milky Way mass functions. All mass functions were re-calculated from the original literature sources and fit with a double power law (Equation~\ref{eq:dbl_pwr_law}) homogeneously across all regions. While there are various differences in how those data were acquired and processed as well as how individual cores and clumps were identified, using the best fit parameters reported by \citet{Rei2006b} will remove any variability introduced in the fitting step. Fig.~\ref{fig:alpha_compare} compares the high and low mass slopes from \citet{Rei2006b} with our fit from \S~\ref{LMC_CMF} for the \gls{lmc}.

The \gls{lmc} break mass is greater than the individual clump masses for all of the Milky Way objects except for W43 and RCW 106 (and greater than all Milky Way break masses which range from \SIrange{0.2 \pm 0.1}{400 \pm 300}{\solarmass}). As a result, it is probably more reasonable to compare the \gls{lmc} low mass slope to the high mass slopes of the Galactic clouds. Fig.~\ref{fig:alpha_compare} shows high mass indices for $\rho$ Oph at \SI{850}{\micro\metre}, NGC 7538 (measured at both \num{450} and \SI{850}{\micro\metre}), and W43 are consistent with the low mass index for the \gls{lmc}. All low mass Galactic indices, except Orion B from \citet{Mot2001}, are systematically shallower than the \gls{lmc} low mass index. The observed consistency between the \gls{lmc} low mass index and the high mass indices from $\rho$ Oph, NGC 7538, and W43 suggests that the shape of the mass function may extend smoothly from the lower mass clumps of the Milky Way regions to the higher mass clumps of the \gls{lmc}. In other words, it appears the mass function slopes at high masses in the Milky Way are extending up to the masses at the bottom of our \gls{lmc} mass function.

In addition, none of the Galactic mass functions had completeness corrections applied and are all likely suffering from incompleteness at their low mass ends. Limiting our comparison to the high mass end of the Galactic mass functions, where incompleteness should have a minimal effect, gives a fairer comparison with our completeness corrected \gls{lmc} mass function.

We note that there are four measurements of Galactic high mass indices which are consistent with our high mass index. Orion B as measured by both \citet{Mot2001} and \citet{Joh2001} is consistent within the uncertainties despite values about \num{20} per cent shallower than ours because the indices have uncertainties of almost \num{25} per cent. The remaining fields whose high mass indices matches ours, M8 and M17, have large enough differences in analysis that we do not believe they present an issue. The cores from M8 were identified by eye rather than algorithmically as well as being fit with an assumed Gaussian profile unlike most of the other clump finding, including our own. For M17, the mass functions were either fit as well or better by a lognormal form so it may not be a fair comparison with our mass function if it is not well characterized by a double power law.

The break masses from the Milky Way regions are all significantly smaller than the best fit break mass for the \gls{lmc}. Galactic break masses range from \SIrange{0.2 \pm 0.1}{400 \pm 300}{\solarmass}. These differences are not surprising as \citet{Rei2010} showed a correlation between distance and break mass using mass functions derived from synthetic observations of a simulated \gls{gmc}. This trend can even be seen among the \num{11} mass functions from \citet{Rei2006b} as the break masses increase in step with decreasing spatial resolution. Our results for the break mass in the \gls{lmc} are reasonable given the much greater distance to the clumps.

\begin{figure}
\centering
\includegraphics{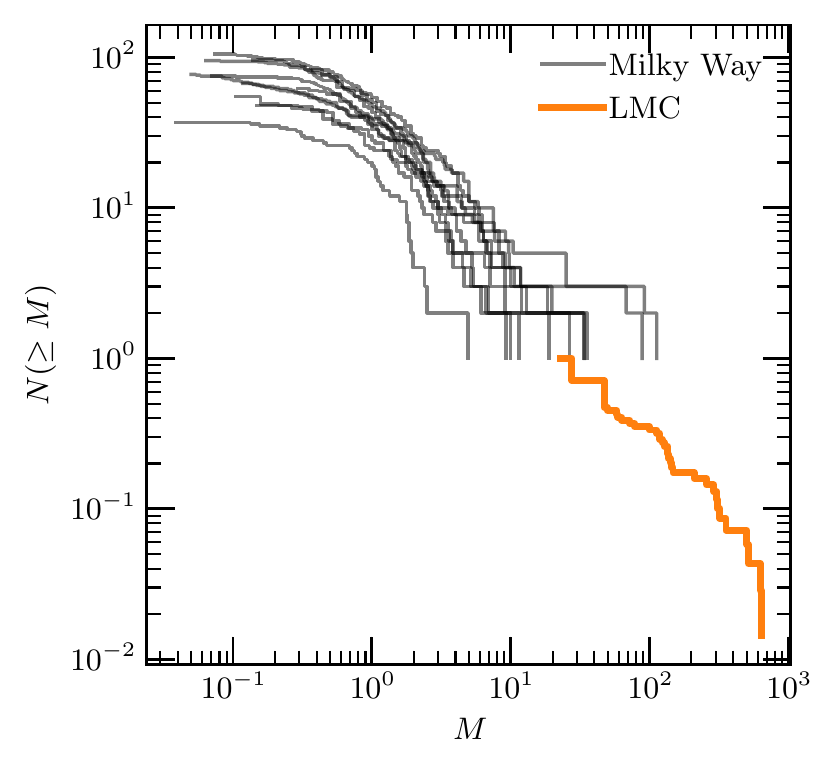}
\caption{All Milky Way mass functions described in \citet{Rei2006b} reproduced from their Fig.~8 in gray. The \gls{lmc} mass function, in orange, has been normalized such that the lowest mass clump is equal to the median of the maximum masses from the Galactic mass functions, and $N(\geq M)$ normalized such that the lowest mass has $N(\geq M)=1$.}
\label{fig:MF_overplot}
\end{figure}

Fig.~\ref{fig:MF_overplot} shows this comparison in another way. All Milky Way mass functions are reproduced from \citet{Rei2006b}, while the \gls{lmc} mass function is scaled such that the lowest mass (\SI{205}{\solarmass}) is equal to the median maximum Milky Way mass (\SI{120}{\solarmass}) and the lowest mass has $\mathrm{N}(\geq\mathrm{M})=1$. Although this scaling is rather arbitrary, it illustrates the potential connection of the mass function shape between the lower masses in the Milky Way and the higher masses in the \gls{lmc}.

Other recent observations have also measured the mass function of cores and clumps in the Milky Way. \citet{Net2009} observed \num{50} square degrees of the Galactic Plane in Vela with the \gls{blast} and fit cold cores (<\SI{14}{\kelvin}) with a power law of \num{-3.22 \pm 0.14} and warmer cores with an index of \num{-1.95 \pm 0.05}. \num{738} cores were extracted from Herschel Gould Belt Survey observations of Aquila by \citet{Kon2015}. They fit their mass function with an index of \num{-2.33 \pm 0.06}, but after applying an age correction to the distribution they fit an index of \num{-2.0 \pm 0.2}. \citet{Pat2017} observed the Cepheus flare region with SCUBA-2 which included four molecular clouds  L1147/L1158, L1174, L1251 and L1228. Individual power law fits gave indices of \num{-1.8 \pm 0.2}, \num{-2.0 \pm 0.2}, \num{-1.8 \pm 0.1} and \num{-2.3 \pm 0.3}, respectively, while fitting all regions together with a double power law gave a low mass index of \num{-1.9 \pm 0.1} and a high mass index of \num{-2.6 \pm 0.3}. These results are in line with our conclusion that the high mass Milky Way mass function slope is similar to the low mass \gls{lmc} slope, except when the entire Cepheus flare region is analyzed together.

Another interesting comparison we can make is to the clump mass function for \gls{dor} derived by \citet{Ind2013} using $^{13}$\gls{co} (2-1) and $^{12}$\gls{co} (2-1) data. Their single power law fit to the differential logarithmic mass distribution of \gls{co} clumps above \SI{500}{\solarmass} resulted in a slope of $\alpha=-1.9\pm0.2$ which is consistent with the low mass index measured from dust masses in this work. Fitting the differential form of the mass function resulted in only about five bins, two of which only have one or two clumps, so it is reasonable that the bins below our break mass would dominate their fit. The similar mass range between our dust estimated masses and their independent \gls{co} estimated masses is evidence for the accuracy of our dust-only map generation and our estimated dust properties.

\citet{Rei2006b} also presented lognormal fits to the Galactic mass functions listed in Fig.~\ref{fig:alpha_compare}. The Milky Way values for $M_{0}$ are all significantly less than the \gls{lmc} value with only RCW 106 coming within ${\sim}2\sigma$, but this is expected as $M_{0}$ is correlated with the mass function median mass like the double power law break mass. We find agreement in the \gls{lmc} $\sigma$ value and six of the Galactic mass functions. Both $\rho$ Oph measurements, M8 and NGC 7538 measured at \SI{850}{\micro\metre} show significantly different best fit values for $\sigma$.

We can also search for a link between the \gls{lmc} clump mass function and the mass distribution of groups of stars as they are first formed. The slope of \num{-2} for young clusters measured by \citet{Lad2003} is quite similar to our low mass slope over a similar mass range. Analogous to how the similar shape of core mass functions and the stellar \gls{imf} has been posited as indicating a constant star formation efficiency, the similarity of the clump and cluster mass functions may result from a constant cluster formation efficiency.

However, dust obscuration challenges observational measurements of the youngest, most deeply embedded star clusters. This hinders our ability to fully constrain the initial mass distribution of clusters as their stars are first born. Cluster formation simulations can bridge the gap between starless molecular clumps and the first star clusters as they emerge from their natal molecular clouds. \cite{Fuj2015} use a hydrodynamical simulation of the parent \gls{gmc} and after about a free-fall time instantaneously remove the gas and continue a solely N-body simulation of the stars formed. They measure mass function indices of \num{-1.73} at \SI{2}{\mega\year} and \num{-1.67} at \SI{10}{\mega\year} for a range of initial \gls{gmc} masses from \SIrange{4e4}{5e6}{\solarmass}. \citet{How2017} aimed to simulate cluster formation with the inclusion of radiative feedback in their hydrodynamical simulations to study its impact on cluster forming efficiency. Starting from \glspl{gmc} of similar masses as \citet{Fuj2015} they found close agreement between their simulated mass function power law index of \num{-1.99 \pm 0.14} and observed embedded clusters. This was after \SI{5}{\mega\year} and for masses from \SIrange{10}{2000}{\solarmass}. For clusters \SI{{>}2000}{\solarmass} they fit a steeper power law of \num{-3.8}. These theoretical results are broadly consistent with the \gls{lmc} clump mass function presented here.

\section{Conclusions} \label{conclusion}
We have presented a clump mass function measured in continuum dust emission from star forming molecular clouds in the \gls{lmc} derived from public archival \gls{alma} observations. \num{32} clumps were observed in the \gls{dor}, N159W, and N159E regions with masses ranging between \SI{205}{\solarmass} and \SI{5740}{\solarmass}. To derive these total masses from the thermal dust emission we presented a technique of combining \num{3.2} and \SI{1.3}{\milli\metre} maps to remove free-free emission contamination and produce dust-only emission maps. We implemented a point source completeness correction to the clump numbers to account for the sensitivity and finite resolution of the observations

From our clump mass function we derived a double power law best fit with parameters $\alpha_{\mathrm{low}} = -1.76^{+0.04}_{-0.05}$, $\alpha_{\mathrm{high}} = -3.3 \pm 0.2$, and $M_{\mathrm{break}} = 2500^{+300}_{-200} \ \mathrm{M}_{\odot}$. Our low mass power law index is similar to the high mass indices of Milky Way regions hosting lower mass cores and clumps. This may indicate an extension of the Milky Way power law to higher masses. The addition of more clumps from more regions in the \gls{lmc} along with mass functions derived from spectral line emission will be necessary to better characterize the mass distribution of these objects in the \gls{lmc} and to investigate its consistency with the Milky Way mass functions.

\section*{Acknowledgements}
We thank the anonymous referee for detailed comments that improved the content of this paper. This paper makes use of the following ALMA data:
\begin{description}
\item ADS/JAO.ALMA\#2011.0.00471.S,
\item ADS/JAO.ALMA\#2012.1.00554.S,
\item ADS/JAO.ALMA\#2012.1.00603.S,
\item ADS/JAO.ALMA\#2013.1.00832.S,
\item ADS/JAO.ALMA\#2013.1.01091.S,
\item ADS/JAO.ALMA\#2013.1.01136.S, and
\item ADS/JAO.ALMA\#2016.1.01533.S.
\end{description} 
ALMA is a partnership of ESO (representing its member states), NSF (USA) and NINS (Japan), together with NRC (Canada), MOST and ASIAA (Taiwan), and KASI (Republic of Korea), in cooperation with the Republic of Chile. The Joint ALMA Observatory is operated by ESO, AUI/NRAO and NAOJ. The National Radio Astronomy Observatory is a facility of the National Science Foundation operated under cooperative agreement by Associated Universities, Inc. This research has made use of NASA's Astrophysics Data System Bibliographic Services. This research made use of Astropy, a community-developed core Python package for Astronomy (Astropy Collaboration, 2013). This research made use of APLpy, an open-source plotting package for Python (Robitaille and Bressert, 2012). This research has made use of the AstroBetter blog and wiki. CDW acknowledges financial support from the Canada Council for the Arts through a Killam Research Fellowship. The research of CDW is supported by grants from the Natural Sciences and Engineering Research Council of Canada and the Canada Research Chairs program.



\bibliographystyle{mnras}
\bibliography{references}



\appendix
\section{Observed, dust-only, and free-free-only maps} \label{all_maps}
Here we present maps for \gls{dor}, N159W, and N159E as observed and also the dust-only and free-free-only emission estimated from our algebraic combination in both Bands 3 and 6. Equation~\ref{eq:b6_dust_only} was used to produce the dust-only maps in Band 6. Free-free-only emission maps in Band 6 were made using Equations~\ref{eq:b6_just_b6} and~\ref{eq:scaled_b3} to solve for the expression
\begin{equation}
\label{eq:b6_ff_only}
S_{\mathrm{ff}} = \frac{S_{\mathrm{B3}} - S_{\mathrm{B6}}\left(\frac{\nu_{\mathrm{B3}}}{\nu_{\mathrm{B6}}}\right)^{(2+\beta)}}{\left(\frac{\nu_{\mathrm{B3}}}{\nu_{\mathrm{B6}}}\right)^{-0.1} - \left(\frac{\nu_{\mathrm{B3}}}{\nu_{\mathrm{B6}}}\right)^{(2+\beta)}}.
\end{equation}
Dust-only and free-free-only emission maps in Band 3 were created by starting instead from
\begin{equation}
\label{eq:b3_just_b3}
S_{\mathrm{B3}} = S_{\mathrm{d}} + S_{\mathrm{ff}}
\end{equation}
and
\begin{equation}
\label{eq:scaled_b6}
S_{\mathrm{B6}} = S_{\mathrm{d}}\left(\frac{\nu_{\mathrm{B6}}}{\nu_{\mathrm{B3}}}\right)^{(2+\beta)} + S_{\mathrm{ff}}\left(\frac{\nu_{\mathrm{B6}}}{\nu_{\mathrm{B3}}}\right)^{-0.1}
\end{equation}
where $S_{\mathrm{d}}$ and $S_{\mathrm{ff}}$ are now the dust-only and free-free-only flux densities at \SI{3.2}{\milli\metre}. We used these to solve for the expressions
\begin{equation}
\label{eq:b3_dust_only}
S_{\mathrm{d}} = S_{\mathrm{B3}} - \frac{S_{\mathrm{B6}} - S_{\mathrm{B3}}\left(\frac{\nu_{\mathrm{B6}}}{\nu_{\mathrm{B3}}}\right)^{(2+\beta)}}{\left(\frac{\nu_{\mathrm{B6}}}{\nu_{\mathrm{B3}}}\right)^{-0.1} - \left(\frac{\nu_{\mathrm{B6}}}{\nu_{\mathrm{B3}}}\right)^{(2+\beta)}}
\end{equation}
and
\begin{equation}
\label{eq:b3_ff_only}
S_{\mathrm{ff}} = \frac{S_{\mathrm{B6}} - S_{\mathrm{B3}}\left(\frac{\nu_{\mathrm{B6}}}{\nu_{\mathrm{B3}}}\right)^{(2+\beta)}}{\left(\frac{\nu_{\mathrm{B6}}}{\nu_{\mathrm{B3}}}\right)^{-0.1} - \left(\frac{\nu_{\mathrm{B6}}}{\nu_{\mathrm{B3}}}\right)^{(2+\beta)}}.
\end{equation}

\begin{figure*}
\centering
\includegraphics{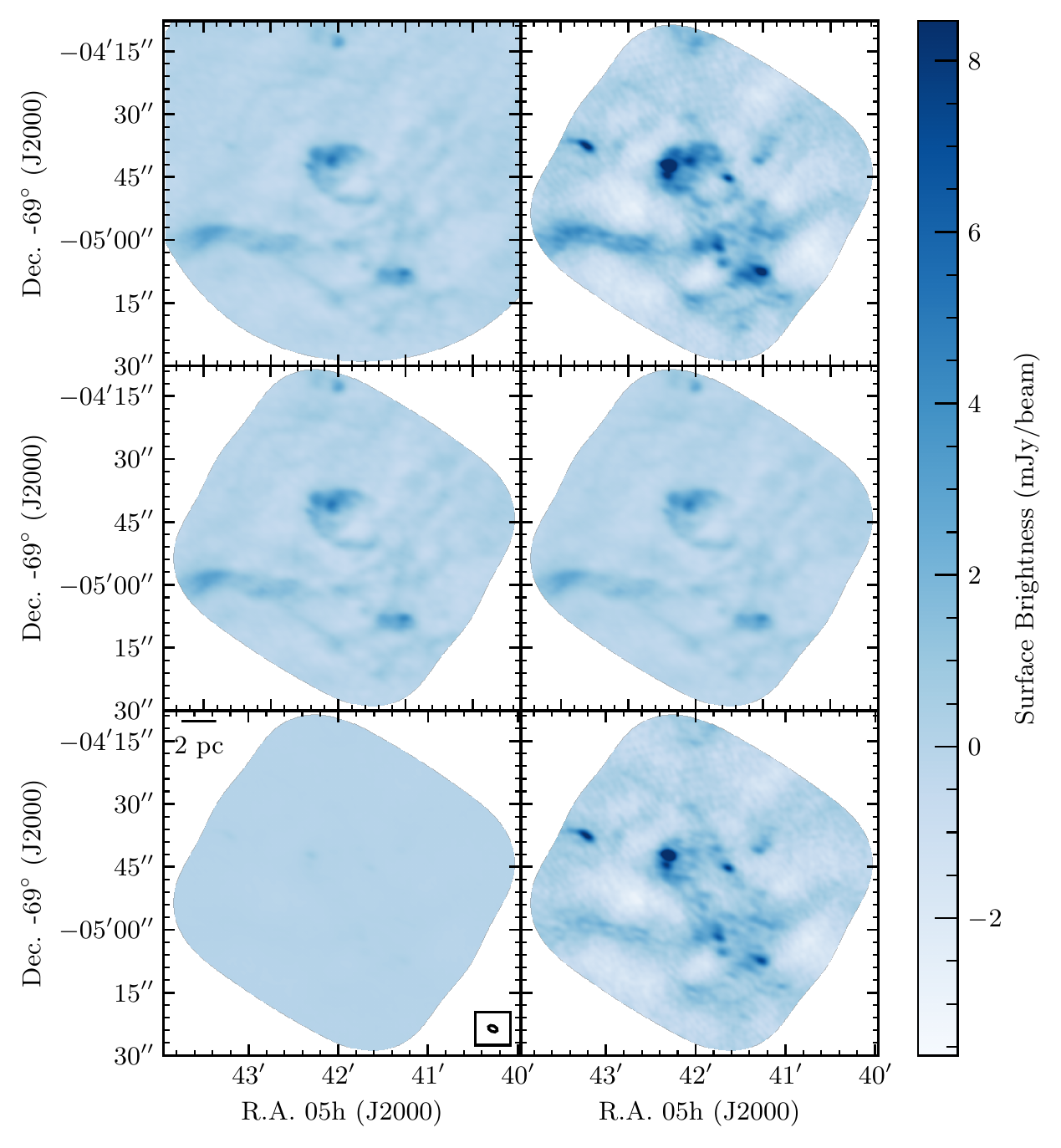}
\caption{Maps of \gls{dor} in \SI{3.2}{\milli\metre} (left column) and \SI{1.3}{\milli\metre} (right column) continuum along with maps of free-free-only (middle row) and dust-only (bottom row) emission at both wavelengths. Spatial and color coordinate axes are matched between each map. The uv coverage of each map has been matched and the synthesized beam shape is represented by the ellipse in the bottom-left panel.}
\label{fig:30Dor_all_maps}
\end{figure*}

\begin{figure*}
\centering
\includegraphics{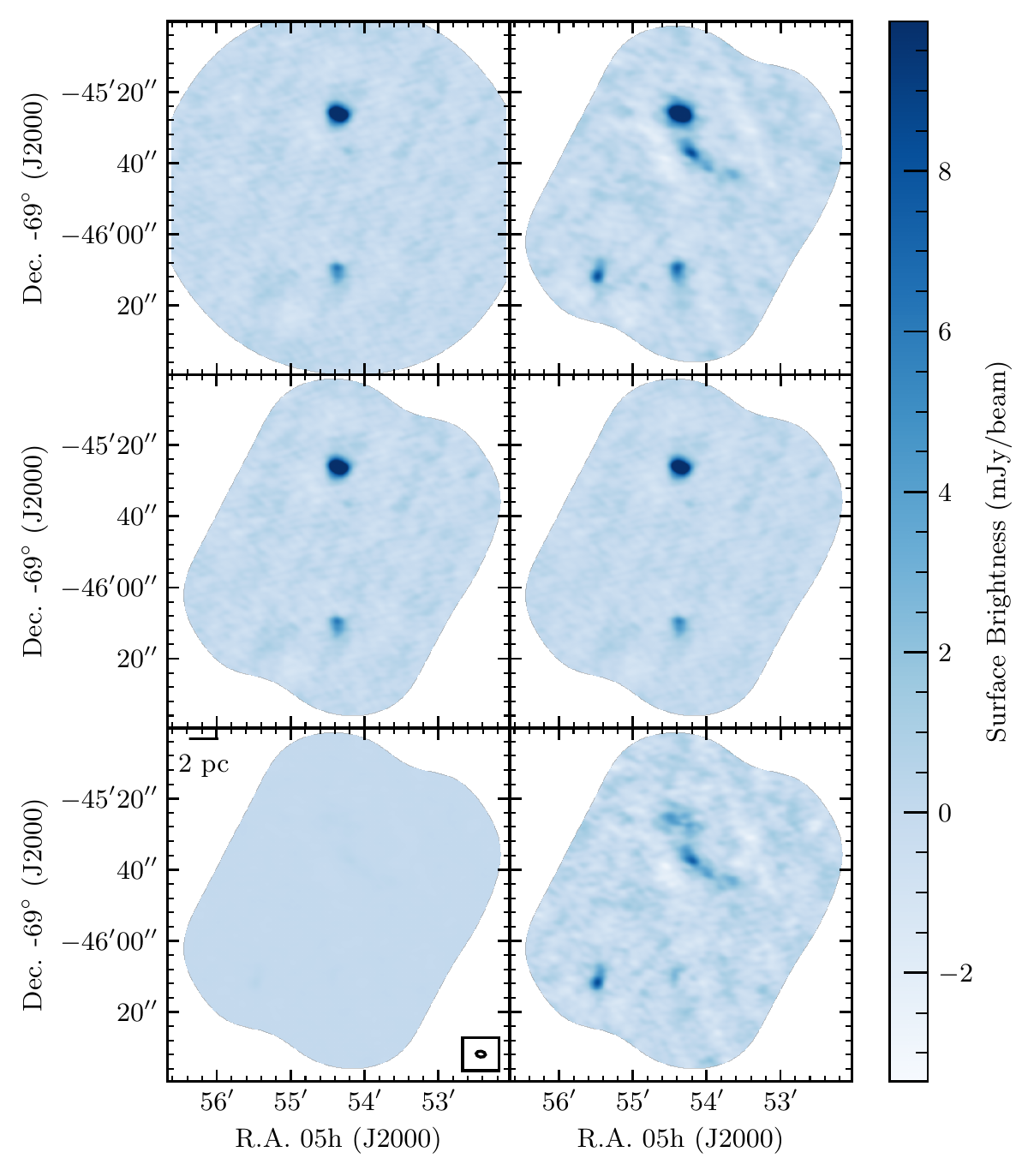}
\caption{Same as Figure~\ref{fig:30Dor_all_maps} but showing N159W observations.}
\label{fig:N159W_all_maps}
\end{figure*}

\begin{figure*}
\centering
\includegraphics{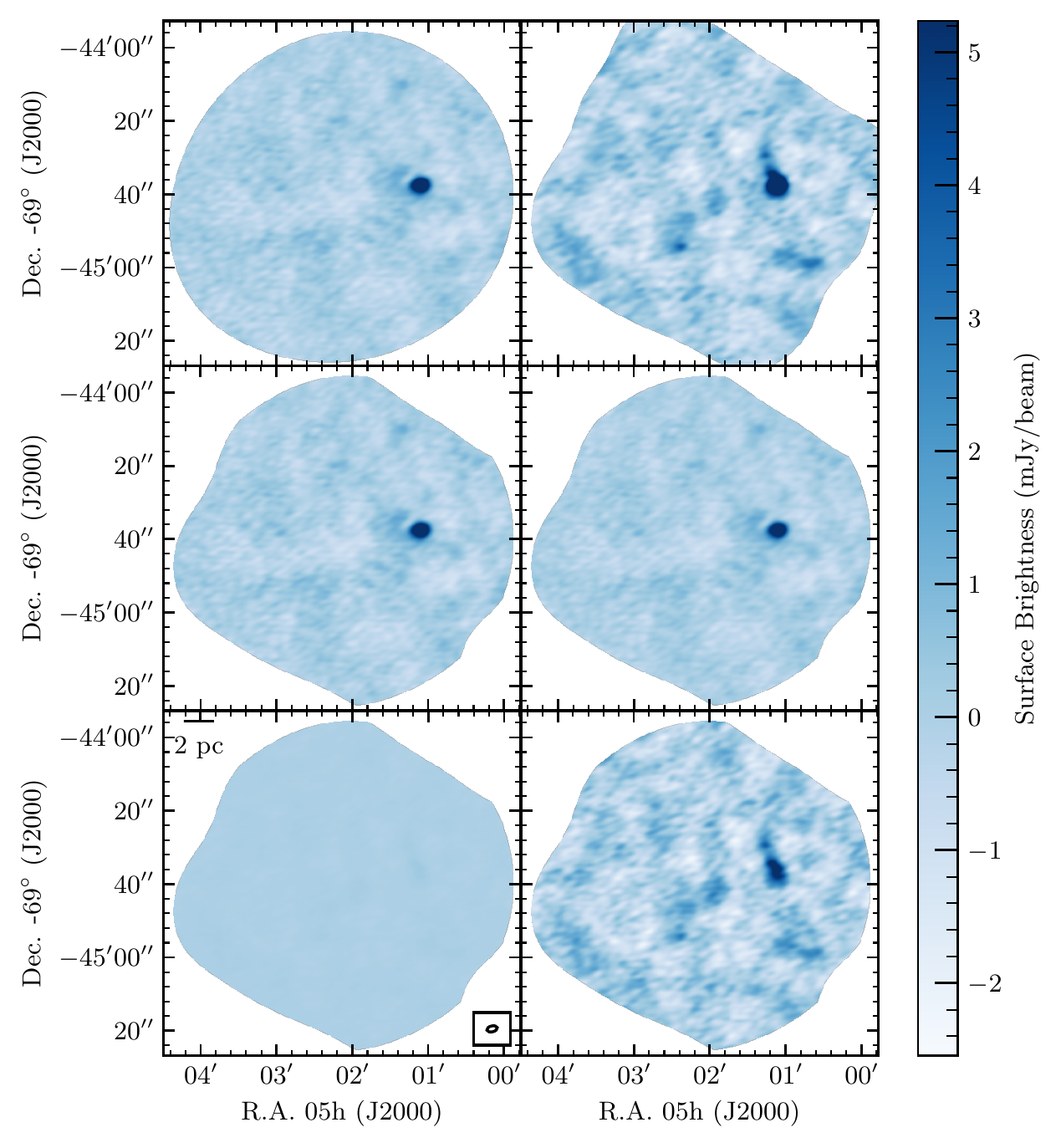}
\caption{Same as Figure~\ref{fig:30Dor_all_maps} but showing N159E observations.}
\label{fig:N159E_all_maps}
\end{figure*}

Figures~\ref{fig:30Dor_all_maps} through~\ref{fig:N159E_all_maps} show all maps produced for \gls{dor}, N159W, and N159E in this work. The top rows show the observed maps, the middle rows the free-free-only emission maps, and the bottom rows the dust-only emission maps. Left columns are maps in Band 3 and right columns are maps in Band 6. All maps are presented with the same spatial and color axes scaling and ranges to facilitate comparisons between the morphologies of the different bands and emission sources.

\gls{dor} exhibits its strongest emission in Band 6 with compact dust sources accounting for most of the flux. N159W and N159E show similar levels of emission in both bands, due to most of the light coming from free-free emission, which slowly varies with frequency. All three fields show the gradual change in free-free brightness between Bands 3 and 6 and the much stronger drop in the dust emission when going from higher to lower frequencies.


\bsp	
\label{lastpage}
\end{document}